\pdfoutput=1

\documentclass[sigconf]{acmart}

\usepackage[T1]{fontenc}

\usepackage[utf8]{inputenc}

\usepackage[linesnumbered,ruled,vlined]{algorithm2e}
\usepackage{algpseudocode}
\usepackage{comment}

\usepackage{graphicx}
\usepackage{enumitem}

\usepackage{amsmath}
\usepackage{url}
\usepackage{bbm}
\usepackage{multirow}
\usepackage{booktabs}
\usepackage{subfig}
\usepackage{blindtext}

\definecolor{darkgreen}{RGB}{0, 100, 0}
\definecolor{purple}{RGB}{128,0,128}
\definecolor{maroon}{RGB}{204,0,0}
\definecolor{navyblue}{RGB}{26,69,149}
\definecolor{Gray}{gray}{0.9}

\newcommand{\user}{u}
\newcommand{\userdecision}[1]{y^{\user,#1}}

\newcommand{\recommendation}{R}
\newcommand{\aidecision}{y^{AI}}
\newcommand{\aiconf}{c^{AI}}

\newcommand{\init}{\text{init}}
\newcommand{\fin}{\text{fin}}

\newcommand{\arcc}{\texttt{ArcC}}
\newcommand{\arco}{\texttt{ArcO}}
\newcommand{\diagc}{\texttt{DiagC}}

\newcommand{\underreliance}{\texttt{Under-Reliance}}
\newcommand{\overreliance}{\texttt{Over-Reliance}}
\newcommand{\switchrate}{\texttt{Switch} \texttt{Rate}}
\newcommand{\totalinapprel}{\texttt{Total} \texttt{Inappropriate} \texttt{Reliance}}
\newcommand{\finalacc}{\texttt{Final} \texttt{Decision} \texttt{Accuracy}}

\copyrightyear{2026}
\acmYear{2026}
\setcopyright{cc}
\setcctype{by}
\acmConference[IUI '26]{31st International Conference on Intelligent User Interfaces}{March 23--26, 2026}{Paphos, Cyprus}
\acmBooktitle{31st International Conference on Intelligent User Interfaces (IUI '26), March 23--26, 2026, Paphos, Cyprus}
\acmPrice{}
\acmDOI{10.1145/3742413.3789136}
\acmISBN{979-8-4007-1984-4/2026/03}

\title{Adjust for Trust: Mitigating Trust-Induced Inappropriate Reliance on AI Assistance}

\author{Tejas Srinivasan}
\affiliation{\institution{University of Southern California}
  \city{Los Angeles}
  \country{USA}}
\email{tejas.srinivasan@usc.edu}

\author{Jesse Thomason}
\affiliation{\institution{University of Southern California}
  \city{Los Angeles}
  \country{USA}}
\email{jessetho@usc.edu}

\begin{document}

\begin{abstract}
User trust biases reliance on AI assistance during decision-making tasks, with overly low and high levels of trust resulting in increased under- and over-reliance, respectively. 
We propose that AI assistants should adapt their behavior through \emph{trust-adaptive interventions} to mitigate such inappropriate reliance. 
For instance, when user trust is low, that providing an AI explanation can elicit more careful consideration of the assistant's advice by the user. 
In two decision-making scenarios---laypeople answering science questions and medical doctors making diagnoses---we find that providing supporting and counter-explanations during moments of low and high trust, respectively, yields up to 38\% reduction in inappropriate reliance and 20\% improvement in decision accuracy. 
We are similarly able to reduce over-reliance by adaptively inserting forced pauses to promote deliberation. 
Our results highlight how AI adaptation to user trust facilitates appropriate reliance, presenting exciting avenues for improving human-AI collaboration.
\end{abstract}


\maketitle

\section{Introduction}
\label{sec:intro}
AI systems are being deployed to assist humans in a wide range of decision-making tasks~\cite{cai2019hello,chiang2023two,che2024integrating}.
AI-assisted decision-making~\cite{lai2023towards} typically consists of an AI system providing a recommendation for a human user's consideration. 
A key factor modulating how users incorporate AI advice is \emph{user trust}, which is the user's belief that the AI will help them achieve their goals in situations characterized by uncertainty and vulnerability~\cite{lee2004trust}. 
Having higher trust makes users more likely to accept the AI's recommendation, all else being equal~\cite{dzindolet2003role}. 
Trust is not a static belief; it evolves as the user interacts with the AI and observes decision outcomes~\cite{dhuliawala2023diachronic}.

\begin{figure}[t]
    \centering
    \includegraphics[width=0.7\linewidth]{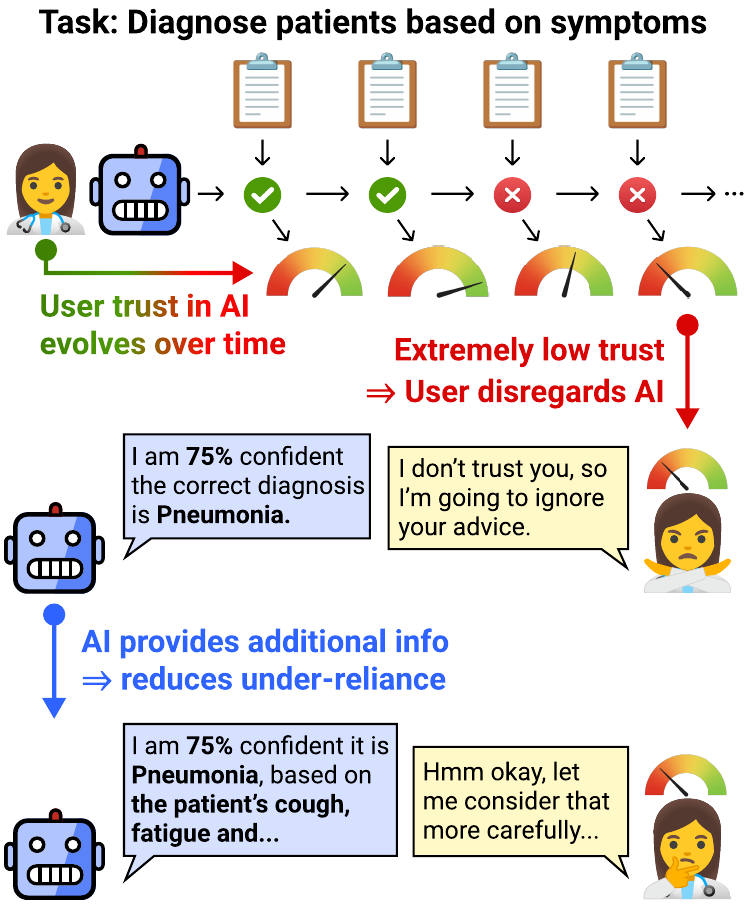}
    \caption{User trust in AI systems evolves over a series of decision-making interactions, impacting how carefully the user considers future AI recommendations. 
    To mitigate the effects of extreme trust and encourage critical reliance, AI systems should adapt their behavior to users' trust levels. 
    }
    \Description{Evolution of user trust in AI assistants, and how AI assistants can adapt their behavior in response to user trust levels.}
    \label{fig:fig1}
\end{figure}

User trust does not always align with the AI assistant's trustworthiness, i.e. its true capability to help the user~\cite{wright2010trust}. 
Miscalibrated trust~\cite{jacovi2021formalizing} may develop due to recency bias, the user's internal biases towards AI, or the assistant's inability to communicate its reasoning or limitations. 
Miscalibrated trust acts as a cognitive bias~\cite{lee2024power} that hinders critical evaluation of AI recommendations, resulting in \emph{inappropriate reliance}~\cite{parasuraman1997humans}. 
For example, we find that medical doctors mistakenly accept 26\% of AI misdiagnoses when their trust is high, compared to 8\% when trust is lower, indicating over-reliance. 
Conversely, when user trust is low, doctors reject correct AI diagnoses 68\% of the time, up from 40\% otherwise, indicating a bias towards under-reliance. 

We posit that AI assistants should adapt their behavior in response to users' trust levels in order to mitigate inappropriate reliance caused by extreme (low or high) trust. 
For instance, when trust is low, the assistant can reduce risk of disuse by providing the user with additional reasoning to support its recommendation (Figure~\ref{fig:fig1}). 
Similarly, when users are too trusting, the assistant can highlight reasons its recommendation may be incorrect via counter-explanations~\cite{si2024large}, or can simply slow down the interaction. 
We hypothesize that strategically deploying \emph{trust-adaptive interventions} will prompt users to engage more carefully with AI advice and yield higher decision-making accuracy.


In this work, we conduct user studies on two decision-making tasks: laypeople answering science questions and medical doctors making clinical diagnoses based on patient symptoms. 
Through controlled studies using a simulated AI assistant, we first validate our premise that user trust levels affect their reliance behavior (Section~\ref{sec:trust_impact}). 
Our analysis reveals that users' decisions to rely on AI assistance are indeed strongly correlated with their trust level at the start of an interaction. 
These findings highlight that trust is an important factor modulating user reliance on AI assistance, and that extreme levels of user trust result in inappropriate reliance of different kinds.

The following experiments (Sections~\ref{sec:mitigate_underreliance},~\ref{sec:mitigate_overreliance},~\ref{sec:mitigate_both}, and~\ref{subsec:deceleration}) investigate the utility of trust-adaptive interventions at mitigating inappropriate reliance caused by extreme levels of user trust. 
We find that adapting AI behavior during at moments of low or high trust by surfacing different kinds of explanations or slowing down interactions is effective at promoting careful deliberation by users and reducing inappropriate reliance. 
Our findings generalize to a setting where users collaborate with a Large Language Model (LLM) assistant (Section \ref{subsec:livellm}), indicating that designing LLMs to be trust-aware presents a promising avenue for improving human-LLM collaboration.

Finally, while our experiments rely on users reporting their trust levels in the AI assistant after each interaction, we also investigate whether these trust levels can be inferred solely from interaction features (Section~\ref{sec:modeling_trust}). 
Our analysis of several heuristic and learning-based methods to estimate trust reveal that using purely interaction-level features is insufficient to accurately predict moments of low and high trust, thus highlighting the challenging nature of trust estimation and the need to consider other human factors.

In summary, our contributions are as follows:
\begin{enumerate}[label=\roman*),nosep]
    \item We first demonstrate that users' trust level at the start of a given interaction affects reliance behavior (\S\ref{sec:trust_impact}), with extreme levels of user trust resulting in worse decision-making. 
    \item Through controlled studies, we find that strategically providing supporting explanations when user trust is low and counter-explanations when trust is high reduces under-reliance and over-reliance, respectively (\S\ref{sec:mitigate_underreliance},~\ref{sec:mitigate_overreliance},~\ref{sec:mitigate_both}). 
    \item We also evaluate the utility of intervening by decelerating the interaction, finding that it helps reduce over-reliance but not under-reliance (\S\ref{subsec:deceleration}). 
    \item We demonstrate that the findings from our controlled studies generalize when applied to a real Large Language Model (LLM) assisting users on decision-making tasks (\S\ref{subsec:livellm}).
    \item Finally, we explore several methods to estimate user trust based on interaction-level features, and find that these trust heuristics and models are insufficient for accurately predicting moments of low and high user trust, presenting open challenges in user trust modeling (\S\ref{sec:modeling_trust}).
\end{enumerate}

Our findings highlight the utility of modeling and adapting to user trust in AI-assisted decision-making, and present exciting avenues for improving human-AI collaboration.

\section{Related Work}
\label{sec:relwork}

We draw from a vast literature on measuring user trust in AI systems and evaluate how decision aids can be used for mitigating inappropriate reliance.

\paragraph{\textbf{Trust in Human-AI Interactions.}} 
Much work has explored the nature of human trust in AI systems~\cite{lai2023towards}, particularly in situations characterized by risk and uncertainty~\cite{jacovi2021formalizing}. 
\citet{lee2004trust} provide the most commonly accepted definition of trust, as a person's attitude that an agent will help them achieve their goals. 
User trust is considered to be calibrated ~\cite{alizadeh2022building} when it aligns with the AI system's true capabilities~\cite{wright2010trust}, thus reducing AI misuse (e.g., over-reliance) and disuse (e.g., under-reliance)~\cite{alizadeh2022building}. 
While trust in AI has typically been attributed to socio-economic and individual factors~\cite{bach2024systematic}, recent work~\cite{dhuliawala2023diachronic, pareek2024trust} examines how trust develops as users interact with AI systems over multiple timesteps. 
Finally, while much work on trust in human-AI interaction is primarily concerned with trust in the AI's ability to improve decision-making performance, trust could also concern other dimensions such as morality, fairness, and transparency~\cite{malle2021multidimensional,jacovi2021formalizing}. 
In this work, we only explore the effect of performance-centric trust on decision-making.

\paragraph{\textbf{Measuring Trust.}} ~\citet{bach2024systematic} identify various mechanisms for measuring user trust, such as questionnaires~\cite{schaffer2019can}, qualitative interviews~\cite{barda2020qualitative}, surveys~\cite{lin2019building}, and point scales~\cite{gulati2019design}. 
In AI-assisted decision-making, trust is often measured by observing user reliance behavior~\cite{yin2019understanding, zhang2020effect}; however, \citet{de2021defining} distinguish reliance, an observable behavior, from trust, a subjective belief. 

Further, even when we are only concerned with performance-centric trust, user trust can be captured from multiple perspectives. 
The Trust in Automation (TiA) questionnaire~\cite{korber2018theoretical} considers six aspects of user trust on system performance capabilities, including reliability/competence, predictability, user familiarity, perceptions of developers' intentions, users' propensity to trust, and overall trust in automation. 
In our user studies, we measure users' perception of AI reliability on a 11-point scale, allowing users to report trust levels with more granularity than a traditional 5-point or 7-point Likert scale.

A final important design decision when measuring performance-centric trust is the granularity at which user trust is captured. Prior works have elicited local, interaction-specific trust levels by asking users to report their confidence in the AI's accuracy for an individual question~\cite{pareek2024trust, dhuliawala2023diachronic}. 
Others adopt a global lens by asking users to report their trust in the AI assistant's general ability rather than in its performance for a specific problem~\cite{korber2018theoretical,schaffer2019can}. 
We elect to elicit global trust scores by asking users to report their belief in the AI's helpfulness after each decision-making problem, and examine how this global trust varies and affects user behavior over the course of multiple interactions.

\begin{figure*}
    \centering
    \includegraphics[width=\linewidth]{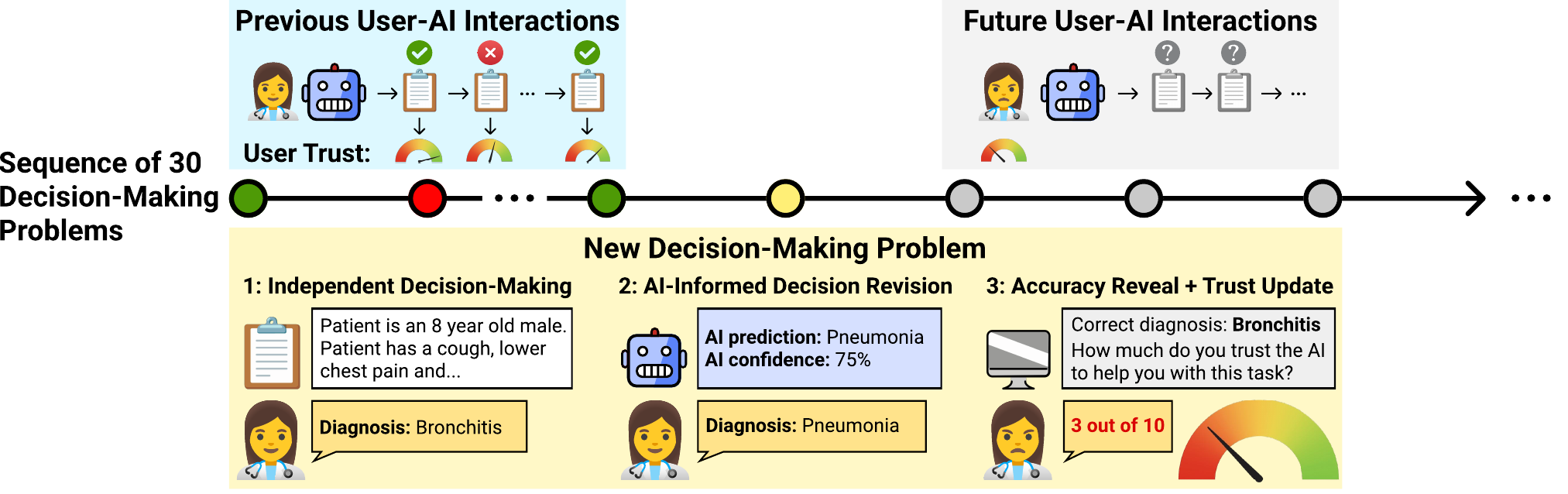}
    \caption{In our user study, each user interacts with an AI system for a sequence of $30$ decision-making problems. In each problem, the user first makes a decision by themselves, and then receives advice from the AI which they use to make a final decision. The user is then told what the correct decision is, and reports their trust in the AI (out of 10).}
    \Description{An illustration of our user study setup, where a user solves each decision-making problem in three phases.}
    \label{fig:decision_making_setup}
\end{figure*}

\paragraph{\textbf{Mitigating Inappropriate Reliance.}} 
Inappropriate reliance, where users mistakenly accept incorrect AI predictions or reject correct ones~\cite{parasuraman1997humans}, is highly undesirable in high-stakes domains, such as healthcare and law~\cite{schemmer2023appropriate}. 
Appropriate reliance can be fostered through various decision aids, such as model confidences~\cite{zhang2020effect}, explanations~\cite{wang2021explanations}, uncertainty expressions~\cite{zhou-etal-2024-relying,kim2024m}, and providing sources~\cite{feng2019can}. 
Cognitive forcing functions~\cite{buccinca2021trust}, which insert friction~\cite{chen2024exploring,inan2025better} and promote deliberation~\cite{park2019slow,rastogi2022deciding}, are effective at mitigating over-reliance. 
We demonstrate how strategically providing these decision aids to users during moments of low or high trust can mitigate trust-induced inappropriate reliance.

\paragraph{\textbf{Adaptive AI Assistance.} }
In the context of AI-assisted decision-making, most work on building adaptive AI assistants has focused on personalizing AI behavior to user attributes, such as their task capabilities~\cite{ma2023should}, their need for cognition~\cite{buccinca2024towards} and their tendency to over-rely~\cite{swaroop2025personalising}. 
In all these works, the user attributes are known or determined before they interact with the AI assistant.
In contrast, we adapt AI behavior to user trust levels, which develop and vary across time as the user interacts with the AI assistant over a sequence of decision-making problems. 

Most similar to our work is~\citet{bhatt2025learning}, which uses an online contextual bandit algorithm to decide which decision support policy to show to users, based on past interactions of the same user with different AI policies. 
This work rests on the assumption that the user's reliance behavior on any given policy will stay stable across time. 
However, we argue that this assumption does not consider how past interactions with a policy will shape users’ trust in that policy, which in turn will alter future reliance behavior.
Instead, we directly condition AI behavior on users' current trust levels.

\section{Preliminaries: Sequential AI-Assisted Decision-Making}
\label{sec:ai_assisted_decision_making}

In this work, we study how user trust impacts reliance on AI advice over a \emph{sequence} of decision-making problems. 
We consider a setting where a human user interacts with an AI assistant on a sequence of $N$ decision-making problems from the same task, such as making medical diagnoses based on symptoms. 
Problems are tuples of input $x$, categorical choices $\mathcal{Y}$, and correct choice $y^*\in\mathcal{Y}$. 
The user solves each problem in three stages: \textbf{independent decision-making} based on their own knowledge, \textbf{decision revision} after viewing the AI advice, and \textbf{trust update} in the AI after observing decision accuracy (Figure~\ref{fig:decision_making_setup}).\footnote{Our setup assumes that both the user and AI can observe the ground-truth decision after each problem.}

\paragraph{1. Independent Decision-Making:} 
For the $i^{th}$ decision-making problem with input $x_i$, user makes a decision $\userdecision{\init}_i \in \mathcal{Y}$. 

\paragraph{2. AI-Informed Decision Revision:} The user views the AI prediction $\aidecision_i$ and confidence $\aiconf_i \in [50\%, 100\%]$, and subsequently makes a final decision $\userdecision{\fin}_i \in \mathcal{Y}$.

\paragraph{3. Trust Update:} After the user makes their final decision $\userdecision{\fin}_i$, they are informed of the accuracy of their decision and the AI prediction $\aidecision_i$. 
Observing this feedback may alter the user's trust in the AI assistant 's ability to help them make better decisions. 
For instance, the AI misleading the user into making a wrong decision is likely to decay trust. 
After viewing the correctness of their decision, users report how much they trust the AI to help them with the decision-making task based on \emph{all} user-AI interactions so far as an integer between 0 and 10.

Our trust operationalization captures the user's evolving belief in the AI's helpfulness, which is likely to influence how the user relies on AI advice in subsequent decision-making problems.

\begin{figure*}[t]
    \centering
    \includegraphics[width=0.98\linewidth]{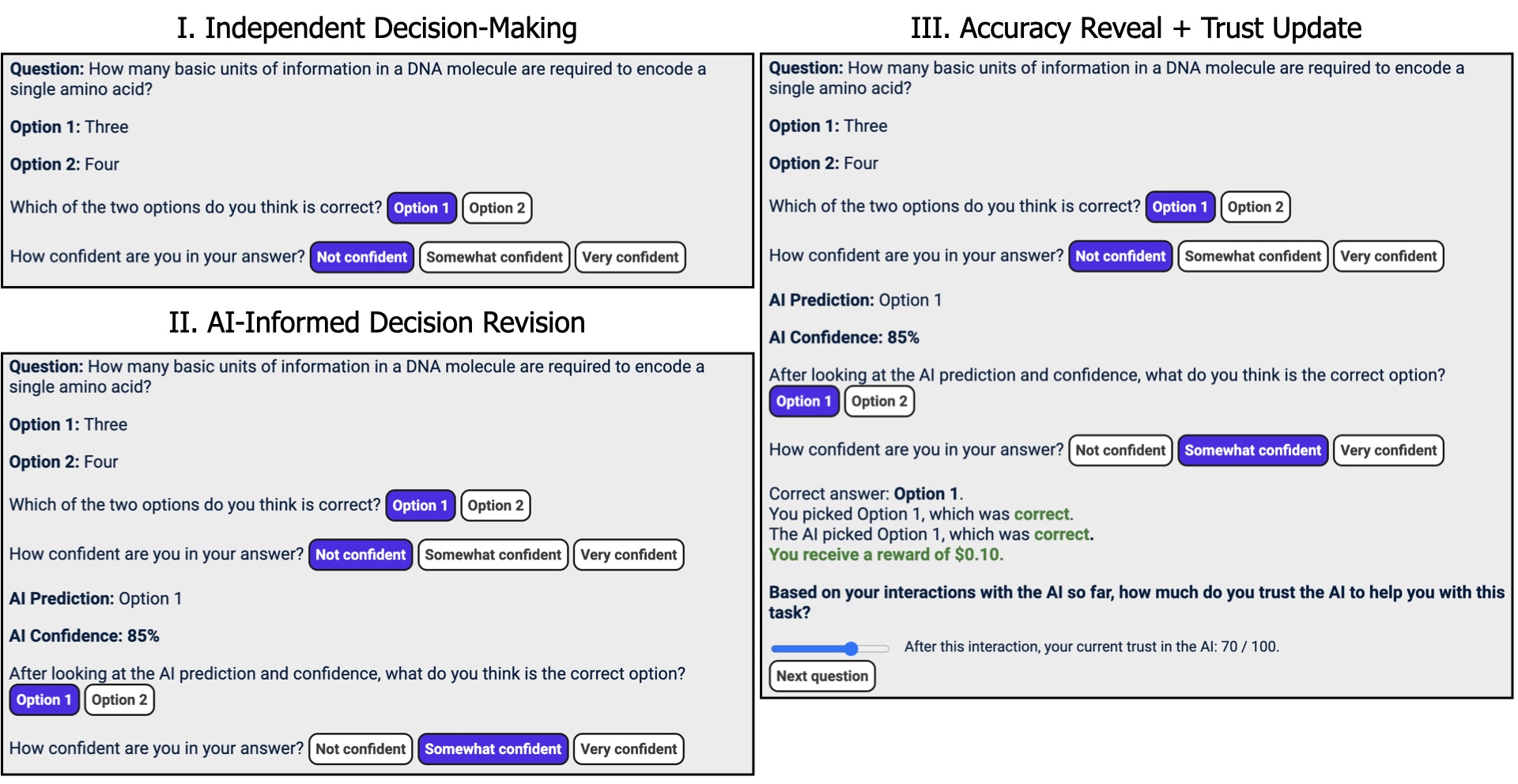}
    \caption{An example of our study interface, where a user solves a decision-making problem from the ARC task in three stages.}
    \Description{An example of our user study interface}
    \label{fig:interface_3step}
\end{figure*}

\section{Experiment Methodology}
\label{subsec:user_study_details}

\subsection{\textbf{Decision-Making Tasks}}
We conduct user studies on two decision-making tasks: \textbf{ARC} and \textbf{Diagnosis}.
Table~\ref{tab:sample_explanations} contains examples of decision-making problems from both tasks. 
Figure~\ref{fig:interface_3step} shows an example of how these decision-making problems are presented to users in our user study interface.

\paragraph{ARC Task.} This task consists of answering science questions. 
Questions are sourced from the ARC dataset~\cite{allenai:arc}, which consists of more than 7000 grade-school science multiple-choice questions written for examinations. 
The authors manually reviewed questions from this dataset and selected questions that were challenging (i.e., the correct answer is not immediately obvious, and at least one option was not obviously incorrect) but still understandable (i.e., did not contain any scientific jargon that laypeople may not be familiar with). 
All questions in the original dataset had four options, but the authors only selected the correct answer and the most plausible incorrect option for the decision-making problem. 
The final filtered set consisted of 39 questions. 

\paragraph{Diagnosis Task.} This task consists of diagnosing patients based on patient age, gender, and symptoms. 
Patient symptoms are sourced from the DDXPlus dataset~\cite{fansi2022ddxplus}, which contains 1.3 million synthetic patients with a differential diagnosis. 
The symptoms consist of an arbitrary number of descriptive statements, such as ``The patient feels pain in their right knee.'' 
We filter down to questions with only 10--15 intake responses so that users do not need to spend a long time reading and understanding the patient case. 
To convert the task into a multiple-choice problem, we select the top three negative conditions from the differential diagnosis as the incorrect options. 
Our final set of problems includes 55 cases corresponding to eleven different conditions.

\subsection{\textbf{Simulated AI Assistance}}

Our user studies use a simulated AI assistant that provides a recommendation ${\recommendation_i = (\aidecision_i, \aiconf_i)}$ for each decision-making problem. 
We experiment with two types of AI assistants that vary in the quality of their confidence estimates: one assistant whose confidence is perfectly calibrated~\cite{guo2017calibration} with prediction accuracy, and one where the AI's confidences are generally higher than their true accuracy (i.e. the AI is overconfident). 

For the calibrated AI assistant, for the $i^{th}$ problem, we first sample a confidence score $\aiconf_i \sim \texttt{Uniform}(0.5, 0.95)$. 
We then decide if the AI prediction $\aidecision_i$ will be the correct decision $y^*_i$ by sampling with probability $\aiconf_i$:
\vspace{-0.5em}
    \[ \aidecision_i =  
        \begin{cases}
        y^*_i & \text{w.p. } \aiconf_i, \\
        \sim \texttt{Uniform}(\mathcal{Y} \setminus \{y^*_i\}) & \text{w.p. } 1-\aiconf_i
        \end{cases}
    \]

For the overconfident AI assistant, we sample the AI confidence $\aiconf_i$ as above, and then sample another parameter $c_i'$ from the triangular distribution $\texttt{Tri}(0.5, \aiconf_i, \aiconf_i)$. 
The AI prediction is sampled as before, but with a correctness probability $c_i'$ lower than the confidence $\aiconf_i$ shown to the user.

This sampling procedure generates AI predictions and confidence scores for each decision-making problem. 

\subsection{\textbf{Participant Recruitment}}

Participants were recruited on the Prolific platform. We recruited participants from the U.K. and U.S.A. who self-identified as fluent in English, and had at least 99\% approval rate on previous studies. 

For the ARC task, we recruit participants on the Prolific platform who have at least an undergraduate degree. 
Users achieve $67\%$ accuracy on this task without AI assistance. 
Users are paid a base amount of \$1.0, with a bonus of \$0.10 per correct final decision. 
For each ARC task setting in our experiments, we recruit 30 users per experimental condition. 

For the Diagnosis task, we conduct studies with professional medical doctors recruited through Prolific. 
These users achieve $74\%$ task accuracy without the AI advice.\footnote{We reject study data from users whose initial decision accuracy is near random chance ($25\%$), since they may have falsified their qualifications on Prolific.} 
Users are paid \$2.0 (since the Diagnosis task takes longer), plus a \$0.10 bonus for every correct final decision. 
Due to the lower number of qualified participants on Prolific, we recruit 20 users per condition in our Diagnosis experiments.

\subsection{\textbf{Task Settings and Sampling Sequences of Decision-Making Problems.}} 
\begin{figure}[t]
    \centering
    \includegraphics[width=\linewidth]{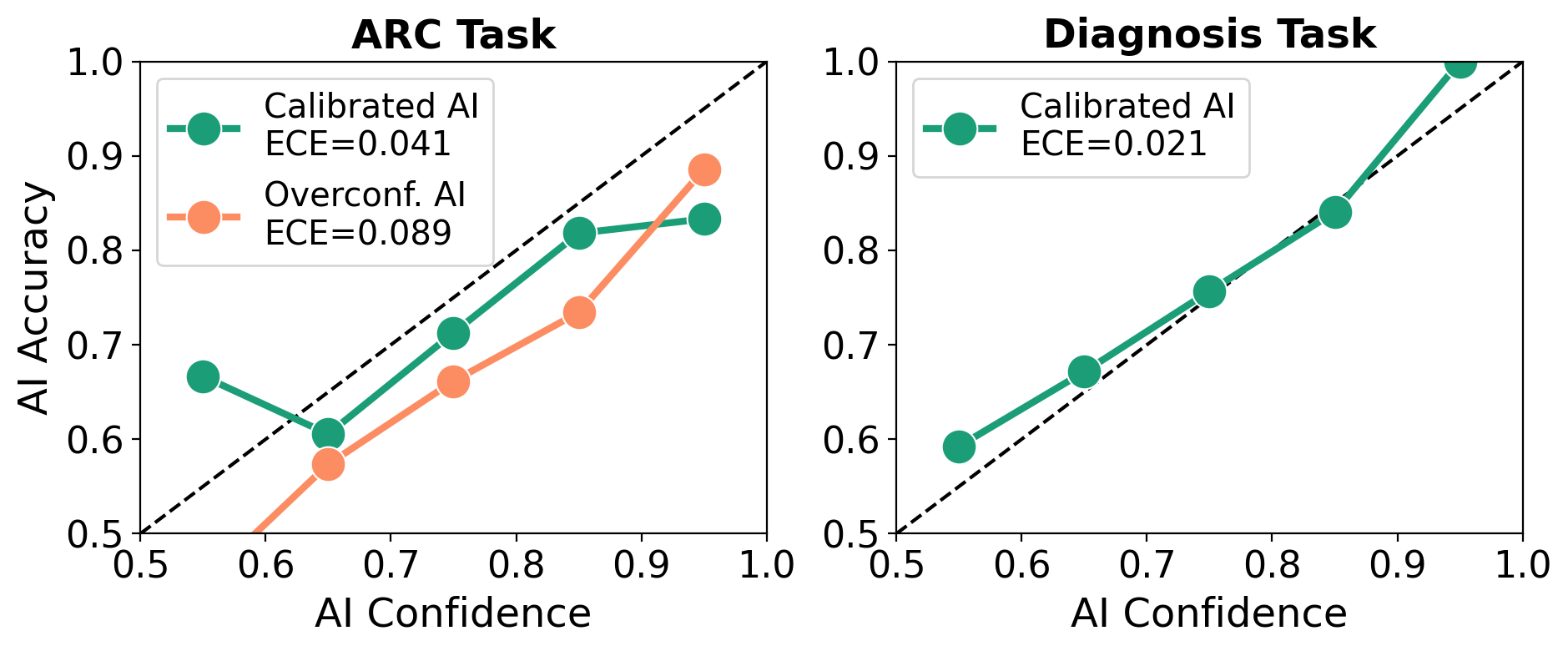}
    \caption{Calibration curves and Expected Calibration Error (ECE) of our simulated AI assistants.}
    \Description{Calibration curves and Expected Calibration Error (ECE) of our simulated AI assistants.}
    \label{fig:ai_calibration}
\end{figure}

We perform user studies with three task settings: the ARC task with a calibrated AI (\arcc), the ARC task with an overconfident AI (\arco), and the Diagnosis task with a calibrated AI (\diagc).\footnote{We do not conduct experiments on the Diagnosis task with an overconfident AI due to being unable to recruit sufficient participants on Prolific.}
For each task setting, we sample 10 sequences $S_i = \{ P^i_1, P^i_2, ..., P^i_{30}\}$ of 30 decision-making problems $P^i_j = \{ (x_j, y^*_j), \recommendation_j \}$, where each problem is also accompanied by an AI recommendation. 
In these sampled problem sequences, the calibrated and overconfident AI assistants have an accuracy of $71\%$ and $64\%$, respectively. 
Figure~\ref{fig:ai_calibration} shows calibration curves for the calibrated and overconfident AI.

Users in each task setting are randomly assigned to a sequence $S_i$ upon starting the study. 
Appendix~\ref{sec:appendix_userstudydetails} contains additional details about the user study setup.

\subsection{Measuring Users' Reliance Behavior} 
\label{subsec:reliance_eval}
We inspect interactions where the user's initial decision differs from the AI prediction, i.e. $\userdecision{\init}_i \neq \aidecision_i$. 
We capture the degree of users' reliance on AI assistance using \textbf{\switchrate}~\cite{yin2019understanding, zhang2020effect}, the fraction of interactions where the user switched their decision to the AI prediction. 
Following~\citet{ma2024you, swaroop2025personalising}, we capture the degree of \emph{appropriate} reliance on AI assistance using two metrics: \textbf{\overreliance} represents the fraction of AI-incorrect interactions where the user switches to the AI's prediction, while \textbf{\underreliance} represents the fraction of AI-correct interactions where the user does not switch to the AI's prediction.

\section{Preliminary Study: How does User Trust Impact Reliance on AI Assistance?}
\label{sec:trust_impact}
In our first experiment, we investigate the correlation between users' self-reported trust levels\footnote{Henceforth, ``user trust'' refers to the user's trust level at the start of an interaction, i.e. the trust score they reported at the end of the previous interaction.} and their reliance behavior in subsequent interactions. 
Previous works have investigated whether users' personality traits, specifically their propensity to trust, are correlated with their reliance on AI assistance~\cite{riedl2022trust,swaroop2024accuracy}; these works have found that these personality traits are not predictive of reliance behavior. 
~\citet{swaroop2025personalising} instead try to determine a user's likelihood of over-relying in future interactions by using probe questions. 
In contrast, we hypothesize that the trust level reported by the user at the end of the $i^{th}$ decision-making problem is correlated with their likelihood of accepting AI recommendations during the $(i+1)^{th}$ user-AI interaction.
\begin{enumerate}[label=\textbf{H\arabic*:}]
    \item When the user's initial decision disagrees with the AI prediction, their likelihood of accepting the AI prediction  (captured by the \switchrate) is correlated with the trust level they reported at the end of the previous decision-making problem.
\end{enumerate}

We further hypothesize that when user trust is either low or high, users make worse decisions about whether to rely on AI recommendations or not in the next user-AI interaction. 
When user trust in the AI assistant is low (i.e., they do not believe the AI assistant can help them with the decision-making task), they are more likely to disregard correct AI recommendations. 
Similarly, when user trust is high, they are more likely to accept incorrect AI recommendations.

\begin{enumerate}[label=\textbf{H\arabic*:}]
    \setcounter{enumi}{1}
    \item Users will have higher \overreliance\ when their trust in the AI assistant at the start of the interaction is high.
    \item Users will have higher \underreliance\ when their trust in the AI assistant at the start of the interaction is low.
\end{enumerate}


\subsection{Experiment Setup}
\label{subsec:initial_trust_experiments}
We conducted user studies across all three task settings (\arcc, \arco\ and \diagc), recruiting 30 participants for each of the ARC task settings and 20 participants for the \diagc setting. 
For each task setting, we aggregate all user-AI interactions across all participants, excluding the very first interaction for each participant since we do not have a user-reported trust level. 
We further exclude all interactions where the user's initial decision $\userdecision{\fin}_i$ matches the AI prediction $\aidecision_i$, since in these interactions users almost always stick with their original decision. 
Finally, we bin the remaining interactions in each task setting by the user's last reported trust level, and compute \switchrate, \underreliance\ and \overreliance\ for each trust bin. 
We compute a weighted Pearson correlation coefficient between user trust levels and the above reliance metrics, where trust bins are weighted by the number of corresponding user-AI interactions.

\begin{figure}[t]
    \centering
    \includegraphics[width=\linewidth]{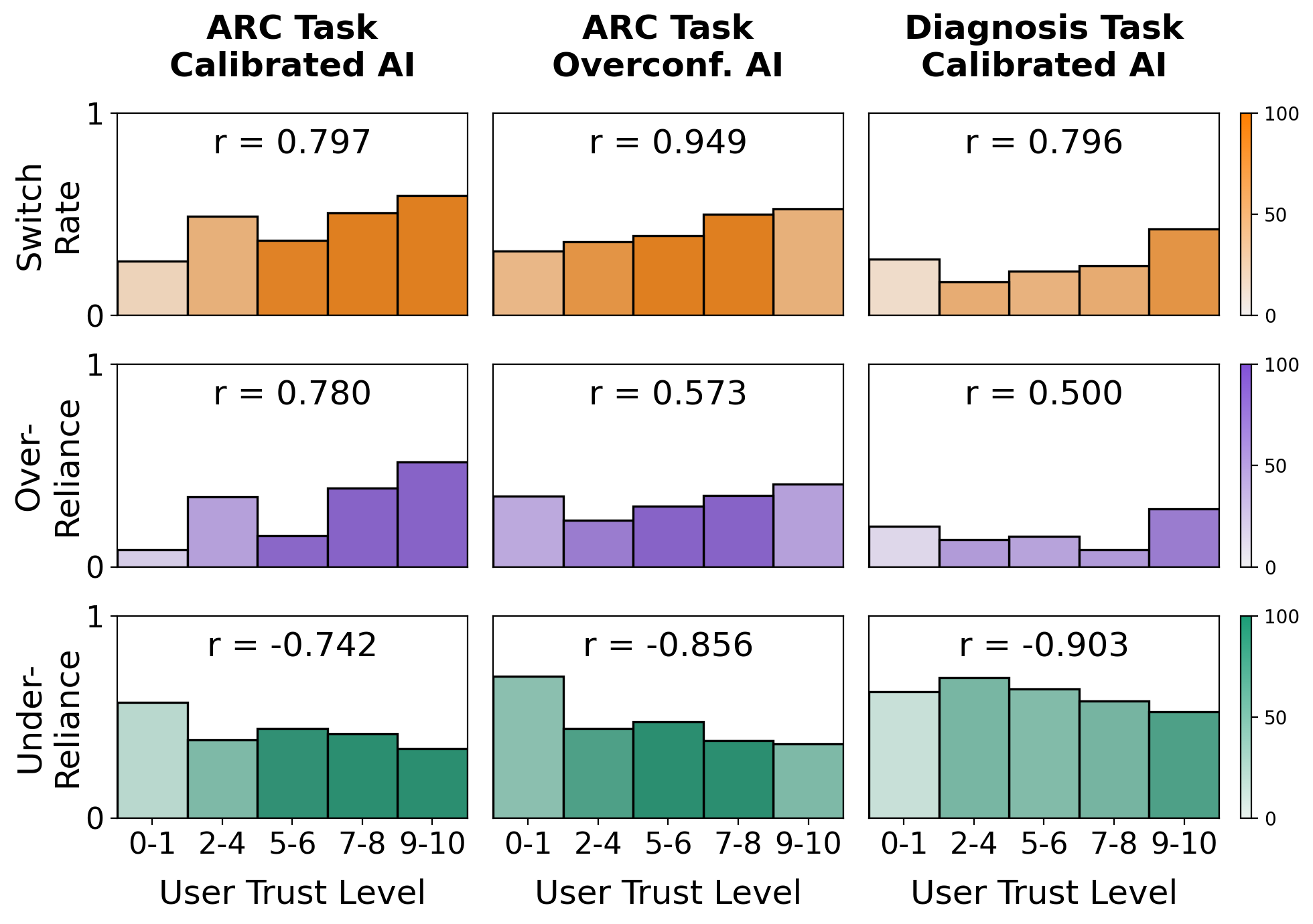}
    \caption{Reliance metrics at different levels of user trust. In each plot, $r$ represents the weighted Pearson correlation coefficient. All correlations are statistically significant, with $p < 0.001$. Bar shades correspond to number of user interactions at each trust level. }
    \Description{Reliance metrics at different levels of user trust.}
    \label{fig:initial_reliance_plots}
\end{figure}

\subsection{Results} 

Figure~\ref{fig:initial_reliance_plots} shows the relationship between user trust and reliance, aggregated across all users in the same task setting. We highlight a few key takeaways:
\paragraph{\textbf{\switchrate\ is strongly correlated with user trust (hypothesis H1):}} Correlations between user trust and \switchrate\ vary from $0.796$--$0.949$, indicating a strong correlation between users' trust levels and their likelihood of accepting AI assistance. These correlations suggest that users' internal trust influences how likely they are to accept AI advice. 
\paragraph{\textbf{Users have higher \overreliance\ when their trust level is high (hypothesis H2):}} User trust has a moderate to strong correlation ($0.500$--$0.780$) with \overreliance. We further observe that \overreliance\ is highest at \emph{high} values of user trust ($9$--$10$). 
\paragraph{\textbf{Users have higher \underreliance\ when their trust is low (hypothesis H3):}}
User trust has a strong negative correlation ($-0.903$ to $-0.742$) with \underreliance. At lower values of trust ($< 5$), users exhibit highest \underreliance.

\vspace{0.5em}
Taken together, these findings suggest that extreme values of user trust act as a cognitive bias for users in subsequent human-AI interactions, resulting in higher inappropriate reliance.

\subsection{Motivating Trust-Adaptive AI Interventions for Mitigating Inappropriate Reliance}
\label{sec:mitigations}

Our findings verify our hypothesis that users' self-reported trust levels (representing the user's belief in the AI's helpfulness at the start of the interaction) is correlated with their reliance behavior in the next interaction. Further, they suggest that when user trust is at extreme levels (either low or high), it acts as a cognitive bias that clouds user judgment when deciding whether to rely on future AI recommendations or not, resulting in inappropriate reliance. 
In a scenario where AI assistants \emph{can} observe a user's trust level after each interaction, 
we investigate whether the AI can counterbalance this trust-induced cognitive bias by adapting its behavior the next time it provides a recommendation to that user. 
For instance, when the user has low trust and is more likely to disregard AI assistance, the assistant can encourage the user to consider the recommendation more closely by providing extra information to support its prediction.
We introduce \emph{trust-adaptive interventions} for reducing inappropriate reliance. 
Trust-adaptive interventions are designed to correct for trust-induced cognitive bias, and are only applied when the user's trust level at the start of an interaction is either above or below a certain threshold. 
Our experiments in the following sections will demonstrate the utility of variance trust-adaptive interventions for mitigating trust-induced inappropriate reliance.

\section{Experiment \#1: Mitigating Under-reliance with Supporting Explanations}
\label{sec:mitigate_underreliance}

AI explanations have been widely studied as a decision aid~\cite{bussone2015role,wang2021explanations,poursabzi2021manipulating}. 
Prior work has shown that natural language explanations supporting the AI prediction cause over-reliance~\cite{si2024large,sieker2024illusion,hashemi-chaleshtori-etal-2024-evaluating}. 
Specifically, it has been shown that when users are provided long-form text explanations, they are prone to believing plausible-sounding explanations without closely examining their accuracy, logical consistency, and reasoning coherence. 
However, when user trust is low, we hypothesize that providing explanations to users will mitigate undesirable disregarding of AI recommendations, while also enabling more critical examination of model explanations. 
\begin{enumerate}[label=\textbf{H\arabic*:}]
    \setcounter{enumi}{3}
    \item Providing natural language supporting explanations when user trust is low will lead to lower \underreliance\ without an equal increase in \overreliance, resulting in an overall increase in users' final decision accuracy compared to not adapting AI behavior in response to user trust.
    \item Providing support explanations adaptively during moments of low trust will yield lower \underreliance\ and higher decision accuracy than providing them during \emph{all} interactions, regardless of users' trust levels.
\end{enumerate}

\subsection{Experiment Setup} 

\subsubsection{Conditions}
We evaluate the utility of adaptive interventions through a between-subjects study.\footnote{We do not conduct a within-subjects study because the same user cannot be subjected to multiple conditions without introducing interaction effects.}
Each user is assigned to one of
three experimental conditions: 
\begin{enumerate}[label=(\arabic*)]
    \item \textbf{No Intervention}: The AI's behavior stays the same (i.e. it always provides a prediction and a numeric confidence score) throughout the user study, regardless of the user's trust level. 
    \item \textbf{Supporting Explanation Always Shown}: The supporting explanation is provided to the user alongside the AI prediction and confidence, regardless of the user's trust level.
    \item \textbf{Trust-Adaptive Supporting Explanations}: The supporting explanation is only provided when the user’s trust lies below a specified threshold. Based on the relationship we observe between reported user trust and reliance (Figure~\ref{fig:initial_reliance_plots}), we select a threshold of 5 out of 10, such that explanations are shown whenever the user's trust level is at 4 or lower. 
\end{enumerate}

Similar to our experiments in Section~\ref{sec:trust_impact}, we conduct studies with 30 users assigned to each experimental condition in the \arcc\ and \arco\ task settings, and 20 users for each condition in the \diagc\ task setting.

\subsubsection{Generating and Providing Supporting Explanations}
We generate supporting explanations for all problems by prompting GPT-4o~\cite{hurst2024gpt} to provide a 3--4 sentence explanation $E^s_i$ for each option $y_i \in \mathcal{Y}$ of each problem $P$ (prompts in Table~\ref{tab:prompts}). 
Explanations were manually reviewed by an author to ensure they entailed the corresponding prediction.
When the intervention was applied during a user-AI interaction, the AI would provide an augmented recommendation $\recommendation'_i = \recommendation_i + E^s_i$. 
To encourage users to read the explanation, they are only allowed to make their final decision 15 seconds after the AI advice is shown.

\subsubsection{Evaluating Interventions.} 
We evaluate the different experimental conditions on which yields the lowest \underreliance\ by users. 
To evaluate whether mitigating under-reliance simultaneously exacerbates over-reliance to the same degree, we also evaluate \textbf{\totalinapprel}, which is the sum of \underreliance\ and \overreliance. 
Finally, \textbf{\finalacc} captures the effect of interventions on users' decision-making performance. 

Similar to Section~\ref{sec:trust_impact}, we compute metrics across all user interactions where the user and AI disagree ($\userdecision{\init}_i \neq \aidecision_i$) and analyze subsets of these interactions based on user trust levels.
Aggregating over interactions rather than users may skew user representation; a user with generally low trust will represent more low trust interactions than other users. 
On the other hand, macro-aggregation by averaging per-user metrics (Appendix~\ref{sec:appendix_macroaggregation}) results in high inter-user variance, since some users have very few interactions meeting the specified criteria. 

We separately evaluated the interventions across interactions where the user's initial decision disagrees with the AI prediction, and across the ``low trust'' subset of interactions. 
For the ``low trust'' interactions, we conduct a clustered bootstrap test~\cite{efron1994introduction} to evaluate significance of our Trust-Adaptive intervention compared to the No Intervention setting.\footnote{We do not conduct a bootstrap test when considering all interactions, because for interactions where user trust is not low, AI behavior is identical in the No Intervention and Trust-Adaptive conditions.} 

\begin{figure*}[t]
    \centering
    \includegraphics[width=\linewidth]{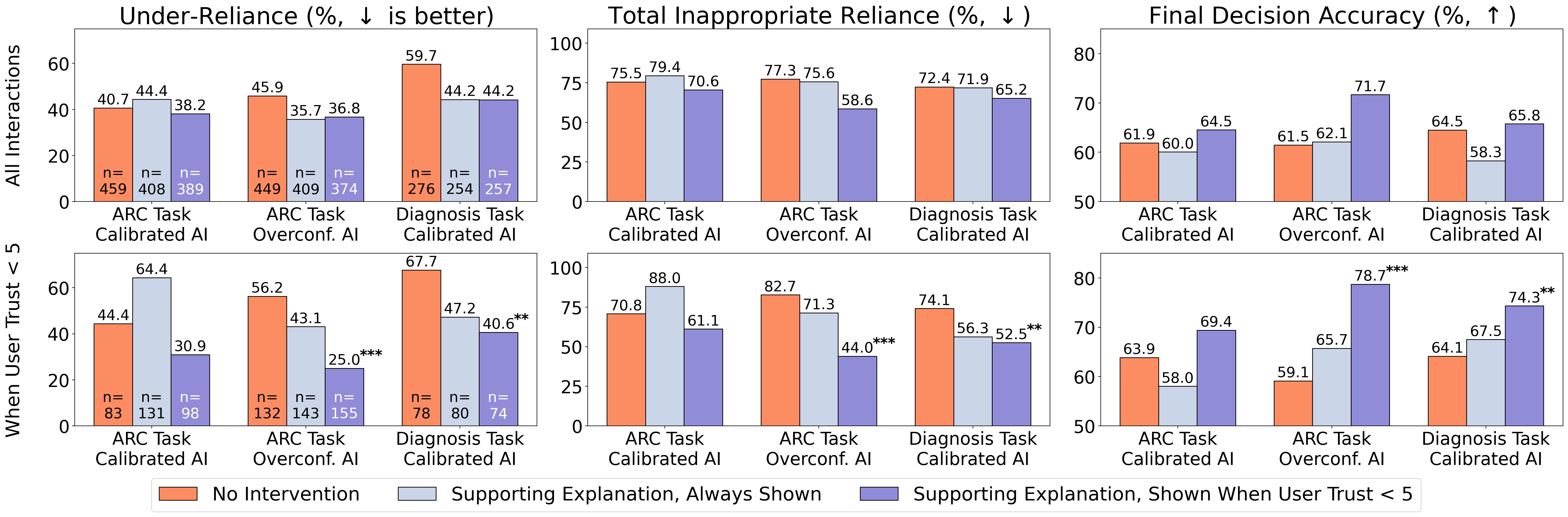}
    \vspace{-2em}
    \caption{Reliance metrics and decision accuracy for users, evaluating the utility of supporting explanations at mitigating under-reliance. 
    $n$ represents the number of user-AI interactions that we aggregate over for the corresponding condition. 
    Showing explanations adaptively reduces \underreliance\ and \totalinapprel\ while boosting \finalacc\ across all task settings, particularly when user trust is low. 
    Asterisks denote significant improvements using a clustered bootstrap test $({^{**}: p < 0.01; ^{***}: p < 0.001)}$.}
    \Description{Reliance metrics and decision accuracy for users, evaluating the utility of supporting explanations at mitigating under-reliance. }
    \label{fig:mitigating_underreliance}
\end{figure*}

\subsection{Results}
Figure~\ref{fig:mitigating_underreliance} shows that \textbf{providing supporting explanations \emph{adaptively} mitigates under-reliance when user trust is low}, especially when AI confidence is miscalibrated.

\paragraph{\textbf{Trust-adaptive explanations reduce under-reliance (hypothesis H4).}}
In the Trust-Adaptive condition, we observe that users exhibit lower \underreliance, lower inappropriate reliance, and higher decision accuracy \emph{across all task settings}. 
When considering interactions where user trust is low, a clustered bootstrap test~\cite{efron1994introduction} reveals that the majority of improvements of the Trust-Adaptive intervention over the No Intervention condition are statistically significant.
Providing explanations when trust is low results in 13--31\% reduction in \underreliance, 9--38\% reduction in \totalinapprel, and 10--19\% improvement in \finalacc.

\paragraph{\textbf{Persistent explanations can hurt, and are not as useful as trust-adaptive explanations (hypothesis H5).}} 
The Supporting Explanation Always condition does not yield similar improvements and worsens decision accuracy in the Diagnosis task. 
In the \arcc\ task setting, showing explanations uniformly \emph{increases} \underreliance\ when trust is low. 
We suspect that showing explanations always rather than adaptively exposes the user to many misleading explanations, resulting in an overall loss of trust in the explanations' trustworthiness. 
In \arco\ and \diagc, explanations reduce inappropriate reliance when trust is low, but not across all user interactions, indicating over-reliance on explanations when trust is not low.

\paragraph{\textbf{Explanations offset AI miscalibration.}} 
The largest improvements occur when the AI system is over-confident, i.e. in the \arco\ task setting, which had the largest number of interactions where user trust was low. 
This finding indicates that explanations are effective for mitigating miscalibrated AI disuse.

\section{Experiment \#2: Mitigating Over-reliance with Counter-Explanations}
\label{sec:mitigate_overreliance}

Similar to how providing supporting explanations is an effective intervention when user trust is low, we investigate whether providing reasons for why the AI prediction might be \emph{incorrect} can counter-balance users' tendency to readily believe AI outputs when their trust is high. 
We hypothesize that alerting users to reasons the AI might potentially incorrect may act as a cognitive forcing function~\cite{buccinca2021trust}, preventing the user from blindly believing AI outputs without due consideration.
Such \emph{counter-explanations}, which present evidence for alternatives apart from the AI prediction, have been shown to reduce over-reliance compared to regular supporting explanations~\cite{cabitza2023let, si2024large, cabitza2024never}. 

\begin{enumerate}[label=\textbf{H\arabic*:}]
    \setcounter{enumi}{5}
    \item Adapting AI behavior when user trust is high by providing counter-explanations to the user will reduce \overreliance\ without an equal increase in \underreliance, resulting in an overall increase in users' final decision accuracy to not adapting AI behavior in response to user trust.
    \item Providing counter-explanations adaptively during moments of low trust will yield lower \overreliance\ and higher decision accuracy than providing them during \emph{all} interactions, regardless of users' trust levels.
\end{enumerate}

\subsection{Experiment Setup}

\subsubsection{Conditions}
Similar to \S\ref{sec:mitigate_underreliance}, each user is assigned to one of
three experimental conditions: 
\begin{enumerate}[label=(\arabic*)]
    \item \textbf{No Intervention}: The AI's behavior stays the same (i.e., it always provides a prediction and a numeric confidence score) throughout the user study, regardless of the user's trust level. 
    \item \textbf{Counter-Explanation Always Shown}: The counter-explanation is provided to the user alongside the AI prediction and confidence score, regardless of the user's trust level.
    \item \textbf{Trust-Adaptive Counter-Explanations}: The supporting explanation is only provided when the user’s trust falls above a specified threshold. We select a threshold trust of 8 out of 10 for applying the trust-adaptive intervention based on the relationship we observe between reported user trust and over-reliance  in Figure~\ref{fig:initial_reliance_plots}. 
\end{enumerate}

\subsubsection{Generating and Providing Counter-Explanations}
Similar to the supporting explanations, we generate natural language counter-explanations (Table~\ref{tab:sample_explanations}) for each option by prompting GPT-4o to list 1--2 reasons why that option might be \emph{incorrect}, while not completely rejecting that option 
(e.g., ``I believe Bronchitis is the correct diagnosis due to\dots, but it is possible that\dots''). 
The counter-explanations frequently include expressions of uncertainty (``could potentially'', ``it may be that''), alternative possibilities, and specific circumstances under which the model's prediction may be incorrect. 

\subsubsection{Evaluating Interventions.} 
Similar to the experiments in \S\ref{sec:mitigate_underreliance}, we evaluate each experimental condition on users' \finalacc\, while also examining whether the two intervention conditions reduces \overreliance\ without increasing the \totalinapprel.

\begin{figure*}[t]
    \centering
    \includegraphics[width=\linewidth]{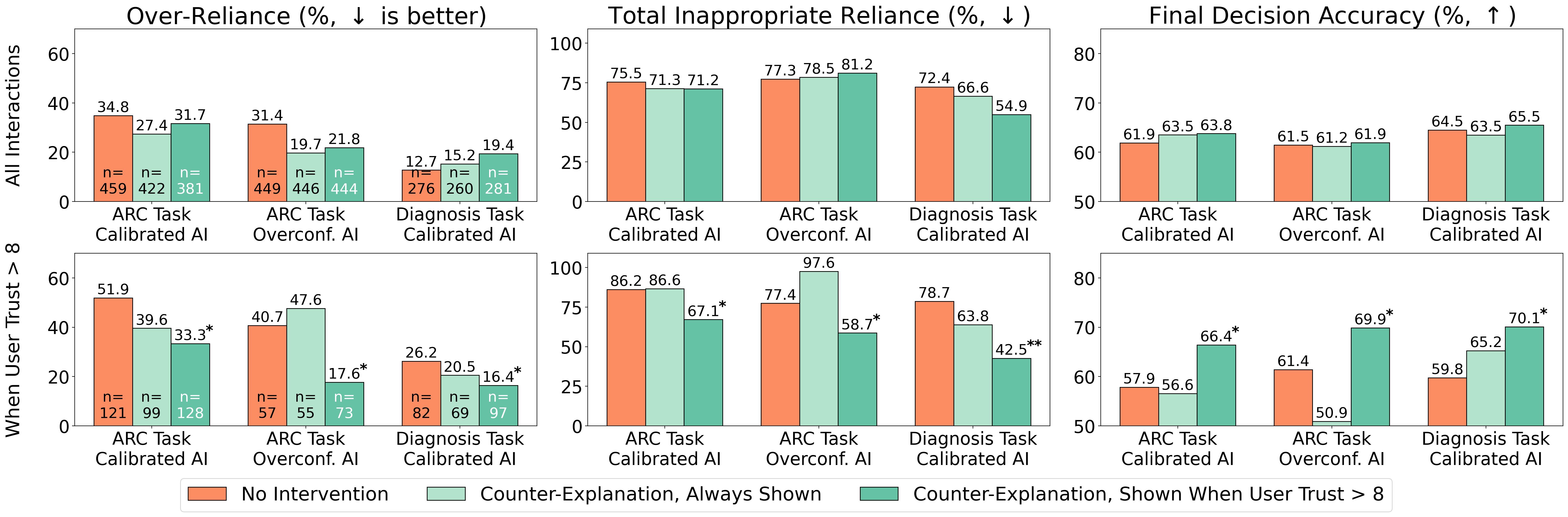}
    \vspace{-2em}
    \caption{Reliance metrics and decision accuracy for users, evaluating the utility of counter-explanations at mitigating over-reliance. Showing counter-explanations adaptively reduces \overreliance\ and \totalinapprel\ while boosting \finalacc\ across almost all settings, particularly when trust is high. 
    Asterisks denote significant improvements using a clustered bootstrap test $(^{*}: p < 0.05)$.}
    \Description{Reliance metrics and decision accuracy for users, evaluating the utility of counter-explanations at mitigating over-reliance.}
    \label{fig:mitigating_overreliance}
\end{figure*}

\subsection{Results}
Figure~\ref{fig:mitigating_overreliance} shows that \textbf{providing counter-explanations \emph{only when user trust is high} reduces over-reliance and improves users' decision-making accuracy.}

\paragraph{\textbf{Trust-adaptive counter-explanations reduce over-reliance (hypothesis H6).}}
Counter-explanations are effective at reducing over-reliance when user trust is high, across \emph{all} task settings, with 10--23\% reduction in \overreliance, 19--36\% reduction in \totalinapprel, and 8--20\% improvement in \finalacc\ (all significant at $p < 0.05$). 
When considering all interactions where the user and AI initially disagree, improvements are also observed in almost all cases but to a lesser extent.

\paragraph{\textbf{Persistent counter-explanations are not as helpful as trust-adaptive ones (hypothesis H7).}} 
The benefits of providing counter-explanations become less pronounced in the Intervention Always condition. 
In the \arco\ setting, users perform uniformly worse than in the No Intervention condition when trust is high. 
These findings indicate that users may be less inclined to closely evaluate counter-explanations when they are always shown.

\section{Experiment \#3: Mitigating Under- and Over-Reliance Simultaneously}
\label{sec:mitigate_both}

Bringing these findings together, we investigate whether showing supporting explanations when trust is low ($<5$) and counter-explanations when trust is high ($>8$) in the same user-AI study yields complementary improvements in decision accuracy and inappropriate reliance. 
\begin{enumerate}[label=\textbf{H\arabic*:}]
    \setcounter{enumi}{7}
    \item Applying both interventions individually (showing supporting explanations when trust is low and counter-explanations when trust is high) will lead to lower \totalinapprel\ and higher \finalacc\ compared to applying each intervention individually.
\end{enumerate}

\subsection{Experiment Setup}

\subsubsection{Conditions} 
We compare the No Intervention baseline and the two earlier trust-adaptive intervention settings applied individually to the setting where the supporting explanation is shown when user trust is low and a counter-explanation is shown when trust is high. 
Due to limited budget for user studies, in this experiment we only study the two task settings corresponding to a calibrated AI assistant (\arcc\ and \diagc).

\subsubsection{Evaluating Interventions} 
Since we are comparing interventions that mitigate both under-reliance and over-reliance, we primarily evaluate conditions based on their \totalinapprel and users' \finalacc. 
We evaluate the four conditions across all interactions, regardless of trust level. 
To understand whether applying both interventions together yields the same results during moments of low and high trust as when they were applied individually, we also look at the subsets of interactions corresponding to low and high user trust. 

\begin{figure*}[t]
    \centering
    \includegraphics[width=\linewidth]{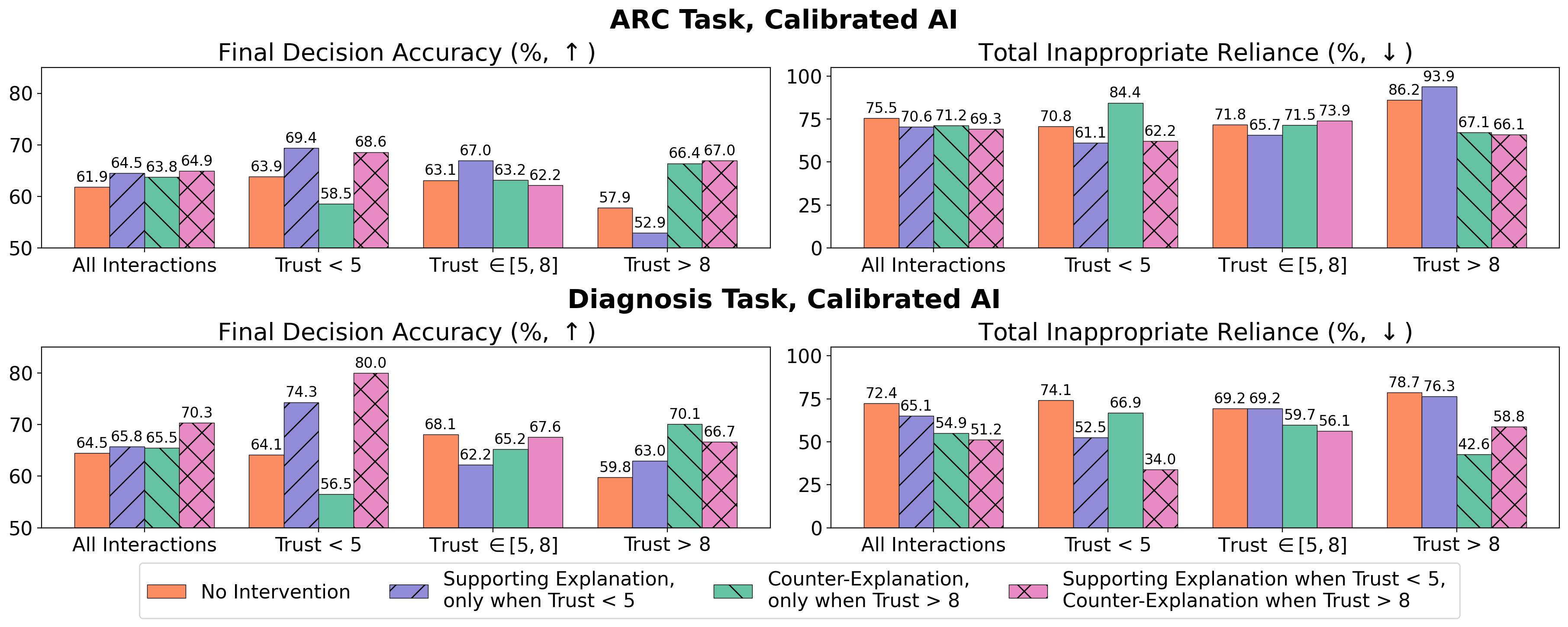}
    \caption{Effect of providing supporting explanations (/) and counter-explanations (\textbackslash), depending on user trust, within the same user session yields complementary benefits in inappropriate reliance and decision-making accuracy.}
    \Description{Effect of providing supporting explanations (/) and counter-explanations (\textbackslash), depending on user trust, within the same user session}
    \label{fig:mitigating_both}
\end{figure*}

\subsection{Results}

Figure~\ref{fig:mitigating_both} shows that \textbf{providing different types of explanations based on user trust yields complementary performance improvements} on both the ARC and Diagnosis tasks. 
When comparing conditions by considering all user-AI interactions in each, the condition where both types of interventions are applied depending on the trust level has the highest \finalacc\ and lowest \totalinapprel\ in both task settings. 
Further, the benefits observed by using supporting explanations during low trust mirror those observed in \S\ref{sec:mitigate_underreliance}, while the benefits of counter-explanation are similar to those in \S\ref{sec:mitigate_overreliance}. 
These findings indicate that when adapting AI behavior to simultaneously combat under- and over-reliance, the two types of interventions do not interfere with each other.

\section{Experiment \#4: Intervention through Deceleration}
\label{subsec:deceleration}

We have demonstrated in Sections~\ref{sec:mitigate_underreliance},~\ref{sec:mitigate_overreliance}, and~\ref{sec:mitigate_both} that supporting and counter-explanations are useful for mitigating under- and over-reliance, respectively. 
However, these need not be the only modes of AI behavior adaptation when user trust is low or high. 
We now investigate the utility of another type of intervention: slowing down the interaction to promote deliberation, without providing additional information to the user. 
Cognitive forcing functions~\cite{park2019slow,buccinca2021trust,rastogi2022deciding} have been widely studied as a tool to encourage users to deliberate and engage in slower, reflective System-2 thinking~\cite{evans2003two}. 

We experiment with two different decelerating interventions. To mitigate under-reliance by decelerating the interaction, we display a ``The AI is thinking\dots'' message for 10 seconds before the AI prediction is revealed to the user. 
We hypothesize that telling the users that the AI is reasoning more deeply (even though no additional reasoning is actually happening) could correct for users' trust being miscalibrated with the AI assistant's trustworthiness.
To mitigate over-reliance, we reveal the AI prediction and ask the user to carefully consider the AI advice; the user must wait 10 seconds before being allowed to make their final decision. 
We hypothesize that these two decelerating interventions can mitigate inappropriate reliance by preventing users from making quick judgments based on their trust levels.
\begin{enumerate}[label=\textbf{H\arabic*:}]
    \setcounter{enumi}{8}
    \item Informing users that the AI is doing additional reasoning when user trust is low can improve users' \finalacc\ and reduce \totalinapprel.
    \item When user trust is high, forcing users to pause before making a decision to rely on AI advice can reduce \totalinapprel\ and improve \finalacc.
\end{enumerate}

\subsection{Experiment Setup}

\subsubsection{Conditions} 
We experiment with the following conditions:
\begin{enumerate}[label=(\arabic*)]
    \item \textbf{No Intervention}: The AI's behavior stays the same throughout the user study, regardless of users' trust level. 
    \item \textbf{Show ``AI is Thinking...'' when Trust is Low}: Whenever the user's trust level is below 5 out of 10, users are shown an ``AI is thinking...'' message for 10 seconds before the AI recommendation is revealed.
    \item \textbf{Add a Forced Pause when Trust is High}: Whenever the user's trust level is above 8 out of 10, after revealing the AI recommendation, users are forced to wait for 10 seconds before making their final decision. 
    \item \textbf{Adaptively showing ``AI is thinking'' or adding forced pause}: Similar to the intervention combination in \S\ref{sec:mitigate_both}, this condition is a combination of the previous two, where AI behavior is adapted when the user's is either low \emph{or} high, but the intervention to be applied depends on the specific trust level. 
\end{enumerate}

\begin{figure*}[t]
    \centering
    \includegraphics[width=\linewidth]{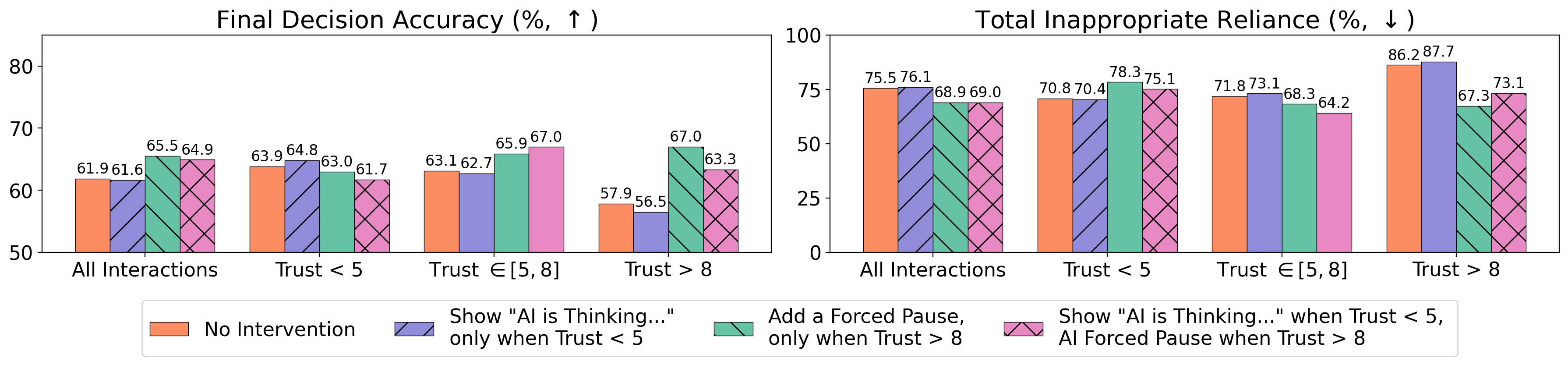}
    \caption{Effect of decelerating interventions on reducing inappropriate reliance and improving decision accuracy.}
    \Description{Effect of decelerating interventions on reducing inappropriate reliance}
    \label{fig:pause_results}
\end{figure*}

\subsection{Results}

Figure~\ref{fig:pause_results} shows the effect of the proposed interventions at mitigating inappropriate reliance in the \arcc\ task. 

\paragraph{\textbf{Showing users an ``AI is thinking'' message does not mitigate under-reliance (hypothesis H9)}} 
We observe that telling users that the AI is thinking is not particularly effective at reducing \totalinapprel, even when trust is low, and users' \finalacc does not improve either. 

\paragraph{\textbf{Adding forced pauses reduces over-reliance (hypothesis H10)}}
We find that forcing users to consider the AI advice closely when trust is high improves \totalinapprel\ and \finalacc. 
Our findings extend those of~\citet{buccinca2021trust}, showing that cognitive forcing is particularly useful for reducing over-reliance when user trust is high.

\section{Experiment \#5: Trust-Adaptive Interventions with an LLM Assistant}
\label{subsec:livellm}

Our studies so far have involved a simulated AI assistant where we control the assistant's accuracy and calibration. 
We now investigate whether our findings generalize when users interact with an actual AI assistant, where we no longer have explicit control of model accuracy and confidence calibration. 
Specifically, we experiment with a Large Language Model (LLM) assistant. 
LLMs are now widely used to help people make better and more informed decisions, including with experts in high-stakes domains~\cite{gumilar2024assessment}, but are prone to cause inappropriate reliance through the use of overconfident~\cite{zhou-etal-2024-relying} and sycophantic~\cite{sharmatowards} language.
\begin{enumerate}[label=\textbf{H\arabic*:}]
    \setcounter{enumi}{10}
    \item If an LLM assistant has access to the user's current trust level, adapting its behavior to provide either supporting or counter-explanations based on user trust level can improve users' \finalacc\ and reduce \totalinapprel.
\end{enumerate}

\subsection{Experiment Setup}

We compare two conditions: the Control condition where there is no intervention, and the Trust-Adaptive condition where the LLM assistance type depends on the user trust level. 
We recruited 30 participants for each of the two conditions, we were assisted by a LLaMA-3.1-8B~\cite{grattafiori2024llama} model on the ARC decision-making task.\footnote{Due to the sparsity of doctors on the Prolific platform, we found it difficult to conduct additional studies on the Diagnosis task, beyond those conducted in the previous experiments.} 

\subsubsection{Generating LLM Recommendations} 
An LLM assistance includes a subset of the following: a prediction, a confidence estimate, a supporting explanation and a counter-explanation. 
We use a simple self-consistency framework~\cite{wangself} to generate all of the above in an efficient batched fashion during the user studies. 
We use chain-of-thought prompting~\cite{wei2022chain} to generate 10 rationale-prediction pairs for each question. The majority predicted answer is selected as the LLM's prediction, with its relative frequency representing the model confidence; we found self-consistency prompting to yield more calibrated estimates than prompting the LLM to output its confidence in its generation. 
In the Trust-Adaptive condition, if user trust is low, the LLM assistant selects one of the chain-of-thought rationales corresponding to the model prediction and provides it to the user as a supporting explanation. 
Similarly, when user trust is high, the LLM assistant would select a rationale corresponding to the other answer (if one exists among the 10 generated pairs) to show as a counter-explanation.

\begin{table}[t]
\small
\begin{tabular}{llrr}
\toprule
User Trust & Metric        & No Intervention & Trust-Adaptive LLM\\
\midrule
\multirow{2}{*}{$<5$ or $>8$} & Decision Acc.                        & 58.4\%          & 67.8\%        \\
& TotalInappRel. & 83.5\%          & 68.2\%       \\
$< 5$ & Under-Reliance                & 65.1\%          & 44.4\%         \\
$> 8$ & Over-Reliance                 & 56.4\%          & 42.4\%         \\
\bottomrule
\end{tabular}
\caption{Effect of providing trust-adaptive supporting and counter-explanations when users are assisted by LLaMA-3.1 to answer questions from the ARC task.}
\label{tab:live_llm}
\end{table}
\subsection{Results}

\autoref{tab:live_llm} compares the effect of both experiment conditions on user reliance outcomes. We observe that \textbf{adapting LLM behavior to user trust improves users' decision accuracy and reduces inappropriate reliance during moments of low and high trust}, mirroring our findings in \S\ref{sec:mitigate_both}, thus validating hypothesis H11. 
Inspecting more closely, we see that under-reliance is reduced when user trust is low and over-reliance is reduced when trust is high. 
These findings indicate that simply having access to a trust signal can allow LLM assistants to improve human-LLM collaboration through trust-aware adaptation.

\section{Analysis: Can We Model User Trust?} 
\label{sec:modeling_trust}
Our decision-making setting assumes that the AI assistant can observe the user's trust level after each interaction. 
However, in real-world settings, the AI assistants may not have access to the trust signal. 
In the absence of this signal, it would be valuable for the AI assistant to be able to predict users' trust levels based on the interaction history.

In this section, we attempt to model user trust directly based on user-AI interaction history, and highlight the difficulties of simple trust heuristics at identifying moments of low and high user trust.

\subsection{Trust Heuristics and Models}
We experiment with several simple heuristics that are computed by looking at interaction outcomes\begin{enumerate}[label=(\roman*)]
    \item \textbf{AIAcc5} estimates trust based on the AI's accuracy in the last 5 interactions. 
    \item \textbf{CapabilityDiff} calculates the difference between the accuracy of the AI's prediction and the user's initial decision, over all interactions so far. 
    \item \textbf{SmoothOutcomes} updates  trust $\tau_t \in [-1, 1]$ after interaction $t$, based on AI prediction correctness $a_t \in \{0,1\}$
        \begin{align*}
            \tau_0 = 0; \tau_t = r \cdot (2 * a_t-1) + (1-r) \cdot \tau_{t-1}
        \end{align*}
        where $r$ is a smoothing parameter.
    \item \textbf{SmoothConfs} is similar to the above, except the update term is weighted by the AI's confidence $\aiconf_t$
        \begin{align*}
            \tau_0 = 0; \tau_t = r \cdot (2 * a_t-1) \cdot \aiconf_t + (1-r) \cdot \tau_{t-1}
        \end{align*}
    \item We train a \textbf{TrustModel}, a linear regression model which estimates the change in user trust after each interaction. 
\end{enumerate}

We use a set of 45 additional user sessions as training data for the model and selecting smoothing parameters, and evaluate on a held-out set of 30 user sessions, totaling 1350 and 900 user-AI interactions in train and test set. 
We evaluate the trust estimation methods on their correlation with user-reported trust levels for the test set interactions, and F1 for predicting low and high trust. 

\begin{table}[t]
    \centering
    \small
    \begin{tabular}{lcccc}
        \toprule
         \shortstack[l]{Trust estimator}& \shortstack[l]{Train corr.} & \shortstack[l]{Test corr.} & \shortstack[l]{High-trust F1} & \shortstack[l]{Low-trust F1} \\
         \midrule
         AIAcc5 & 0.388 & 0.345 & 0.345 & 0.458 \\    
         CapabilityDiff & 0.185 & 0.248 & 0.382 & 0.500 \\ 
         SmoothOutcomes & 0.466 & 0.434 & 0.495 & 0.447  \\ 
         SmoothConfs & 0.483 & 0.430 & 0.486 &  0.423  \\ 
         TrustEffectModel & 0.478 & 0.509 & 0.392 & 0.498  \\
         \bottomrule
    \end{tabular}
    \caption{Correlation of trust estimation methods with user-reported trust levels, and F1 for detecting moments of low and high trust.}
    \label{tab:trust_modeling}
\end{table}


\subsection{Results} 
Table~\ref{tab:trust_modeling} shows correlation between trust scores estimated by the various heuristics and the ground-truth user-reported trust levels, on both train and tests, along with the F1 for detecting moments of low and high user trust. 
We observe that all of the above methods achieve only moderate correlation with user-reported trust, and achieve very low F1 ($\leq 0.5$) at detecting both low and high trust instances. 
These results point to the challenging nature of modeling user trust, and the inadequacy of surface-level interaction features. 
Future work can model trust using more sophisticated interaction features~\cite{fang2013trust}.

\section{Discussion}
\label{sec:discussion}
Our results highlight the promise of trust-adaptive interventions based on experiments in a controlled setting. 
We highlight some considerations when designing trust-adaptive interventions for real-world decision-making scenarios.

\paragraph{\textbf{Implications for NLP systems.}} 
While our findings highlight the benefits of adapting AI behavior to user trust and prior user-AI interactions, LLMs currently generate responses without considering user trust. Our findings suggest LLM assistants can be improved by: 1) training or prompting LLMs to estimate user trust based on conversational history, 2) training LLMs to adaptively decide when and how to insert trust-adaptive interventions in their generations to support reliable user decision-making.

Further, while free-text LLM explanations have been widely studied~\cite{rajani2019explain, wiegreffe2022reframing}, it is unclear whether they are actually useful to users~\cite{joshi2023machine, si2024large, hashemi-chaleshtori-etal-2024-evaluating}. 
Our work adds to this discourse by finding that LLM-generated explanations are useful for mitigating inappropriate reliance when user trust is low or high.

\paragraph{\textbf{Applying trust-adaptive interventions.}} 
A key assumption of our setup does not hold in most real-world settings: that both parties have access to real-time feedback about decision accuracy. 
Environment feedback in response to user decisions will provide a sparse signal in some contexts. 
Users may still internally update their trust in the AI based on their confidence in their own decision and their (potentially incorrect) perception of their expertise in comparison to the AI assistant's ability.
Additionally, some interventions may not be suitable for certain tasks; for example, LLM-generated explanation frequently known to hallucinate details, which may be undesirable in high-stakes applications.

\paragraph{\textbf{Is all inappropriate reliance \emph{equally} bad?}}
Our formulation of \totalinapprel\ treats both under- and over-reliance as equally undesirable.
However, a company developing an AI assistant for clinicians may be more wary of clinicians over-relying on their assistant, which may leave them liable. 
The relative importance of mitigating under- and over-reliance can be quantified through a balancing utility function.
Such a utility function can also be used as a reward signal for optimizing an intervention policy to signal when an intervention should be applied.

\paragraph{\textbf{Should AI assistants be selectively transparent? }}
The research community has historically regarded transparency about AI behavior and reasoning through model explanations as essential for responsible and accountable deployment of AI systems~\cite{schmidt2020transparency,liao2023ai,felzmann2020towards}. 
However, many previous works have found that explanations can often be at odds with optimal decision-making performance by causing users to over-rely on AI assistance~\cite{zhang2020effect,bansal2021does}. There is therefore a need to design explainable AI systems that balance transparency with appropriate reliance~\cite{de2024we}. 
However, many previous works have found that explanations can often be at odds with optimal decision-making performance by causing users to over-rely on AI assistance~\cite{zhang2020effect,bansal2021does}. There is therefore a need to design explainable AI systems that balance transparency with the need to encourage appropriate reliance~\cite{de2024we}. 
To this end, the community has investigated the utility of \emph{selectively transparent} systems~\cite{muralidhar2024effect,MURALIDHAR2025103591,franccois2022transparent}, where explanation details may be disclosed progressively~\cite{muralidhar2024effect} or when demanded by users~\cite{MURALIDHAR2025103591}. 
Our work extends this line of research by conditioning when and how to be transparent about AI reasoning on users' trust levels. 
We find that using trust as a cue to modulate AI transparency improves users' decision-making and reduces inappropriate reliance. 
However, selective transparency may not be appropriate for all contexts; we urge AI system deployers to tailor the degree of AI transparency based on the specific deployment context and the transparency-performance trade-off for their application.

\paragraph{\textbf{Limitations.}} 
The majority of our user studies were performed using a simulated AI assistant, rather than a real system such as a Large Language Model. 
Similar to~\citet{buccinca2021trust,dhuliawala2023diachronic}, we use a simulated AI for the controllability of AI accuracy and confidence calibration. 
However, a simulated AI may be unrealistic compared to real AI systems that people use. 
For instance, since the answer correctness is decided independently for each problem, our AI assistant may provide contradicting predictions for near-identical problems.

Furthermore, the behavior of Prolific participants interacting with an AI in a user study may differ from users of a live system in a real-world scenario~\cite{mcgrath1995methodology}, particularly due to misalignment of motivations~\cite{deci1999meta}. 
We test for the effect of two confounding variables: trust reporting and user experience with AI (Appendix~\ref{sec:appendix_confoundingvariables}), finding that they do not affect user reliance behavior, but there may be other confounding variables we have not yet considered.

We conducted all our user studies with participants from the U.K. and U.S.A. who were fluent in English. 
Users from other countries and cultures may have different attitudes towards AI systems, and thus behave differently.

Finally, our findings would be more reliable if we could obtain data from more participants, more tasks and more interventions, made difficult due to financial restrictions and the dearth of professional doctors on the Prolific platform.

\paragraph{\textbf{Ethical Considerations.}} 
While modeling user trust can allow AI systems to help users overcome their cognitive biases, care must be taken that such user modeling is not taken advantage of to mislead or manipulate users. 
We emphasize that our trust-adaptive interventions are not intended to force or manipulate users into behaving a certain way, but to recognize when the user's trust may be hindering their reasoning.
We do not condone use of such user modeling to manipulate users by misrepresenting the AI's beliefs or causing the user any distress. 

People designing AI assistants with trust-adaptive interventions should ensure the interventions comply with the local ethical standards of the intended users. 
Further, we encourage promoting transparency and accountability by disclosing to users that the AI assistant is modeling user trust.

\section{Conclusion and Future Directions}
\label{sec:conclusions}
We explore the utility of \emph{trust-adaptive interventions} for mitigating inappropriate reliance when users have low or high trust in AI assistants. 
We demonstrate that low and high levels of trust result in increased inappropriate reliance on AI recommendations for laypeople answering science questions and for doctors making medical diagnoses.
We conduct controlled between-subjects studies and find that adaptively providing supporting explanations during low trust and counter-explanations during high trust reduces inappropriate reliance and improves users' decision accuracy. 
These findings generalize to decelerating interventions; forcing users to pause and deliberate before making their final decision helps reduce over-reliance.

Our findings present an initial exploration into adapting AI behaviors based on user trust levels.
We adopted a simple thresholding criterion for deciding when to intervene, but more sophisticated criteria that also account for user and AI confidence may have potential. 
Further, rather than looking at whether user trust is too high or low, we can consider whether the trust is calibrated with the assistant's trustworthiness. 
For example, high user trust may not be as undesirable when the AI is significantly more accurate than the user on the task. 
We hope our findings inspire the community to more closely consider the effect of user trust in user-AI interactions, and to invest more resources into the study of modeling user trust.



\begin{acks}
    We thank Rishi Vanukuru, Gregory Yauney, Sayan Ghosh and Jesse Zhang for their invaluable discussions and brainstorming, as well as Malihe Alikhani, Heather Culbertson, Suvodip Dey, Vardhan Dongre, Mert \.Inan, Sunnie Kim, Chaitanya Malaviya,  Ana Marasovi\'c, Weiyan Shi, Anthony Sicilia, Swabha Swayamdipta, Dilek Hakkani-T\"ur and G\"okhan T\"ur for their valuable feedback. 
    We also thank~\citet{dhuliawala2023diachronic} for open-sourcing their user study interface, and in particular Vil\'em Zouhar for his assistance with setting up the user study and data logging. 
    We are also very grateful to Aisha Baptiste for swiftly adding money to our Prolific account whenever required. 
    
    Tejas Srinivasan was supported by the Amazon ML Fellowship from the USC-Amazon Center for Secure and Trusted Machine Learning. 
    This work was supported in part by a grant from the DARPA Friction and Accountability in Conversational Transactions (FACT) AI Exploration (Award HR00112490376) and in part by a grant from USC-Capital One Center for Responsible AI and Decision Making in Finance (CREDIF) to the first author.
    The opinions and findings expressed in this work do not necessarily reflect those of USC, Amazon, Capital One, or DARPA.
\end{acks}

\bibliographystyle{ACM-Reference-Format}
\bibliography{custom}

\appendix

\begin{figure*}[t]
    \centering
    \includegraphics[width=\linewidth]{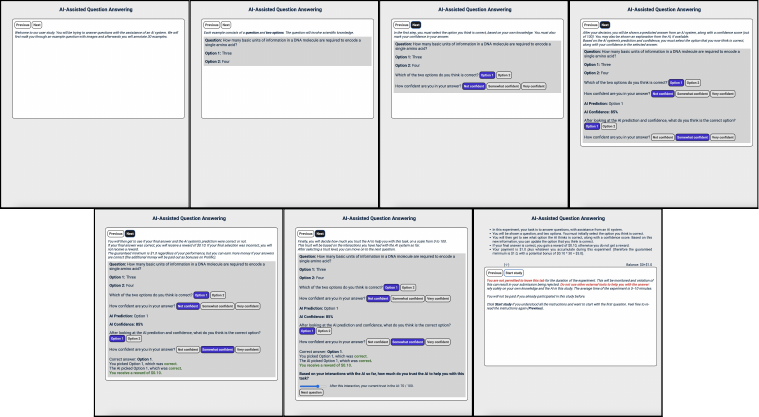}
    \caption{Screenshots of task instructions shown to users.}
    \Description{Screenshots of task instructions shown to users.}
    \label{fig:interface}
\end{figure*}

\newpage

\section{Task Details}
\label{sec:appendix_taskdetails}
We conduct user studies on two decision-making tasks: the \textbf{ARC} task and the \textbf{Diagnosis} task.

\paragraph{ARC Task.} This task consists of answering science questions. 
Questions are sourced from the ARC dataset~\cite{allenai:arc}, which consists of more than 7000 grade-school science multiple-choice questions written for examinations. 
The authors manually reviewed questions from this dataset and selected questions that were challenging (i.e. the correct answer is not immediately obvious, and at least one option was not obviously incorrect) but still understandable (did not contain any scientific jargon that laypeople may not be familiar with). 
All questions in the original dataset had four options, but the author only selected the correct answer and the most plausible incorrect option for the decision-making problem. 
The final filtered set consisted of 39 questions. 
For each of the 10 problem sequences that users solve, we sample 30 of the 39 questions without repetition.

\paragraph{Diagnosis Task.} This task consists of diagnosing patients based on patient symptoms. 
Patient symptoms are sourced from the DDXPlus dataset~\cite{fansi2022ddxplus}, which contains 1.3 million synthetic patients with a differential diagnosis. 
The symptoms are presented as either binary, categorical or continuous variables, but each symptom has a corresponding patient intake question and answer in English, which we translate into a descriptive third-person statement using GPT-4o.. 
For example, the intake question ``Do you feel pain somewhere?'' and patient response ``Knee (R)'' is translated into ``The patient feels pain in their right knee.''. 
We filter down to questions with only 10--15 intake responses so that users do not need to spend a long time understanding the problem. 
To convert the task into a multiple-choice problem, we select the top three negative conditions from the differential diagnosis as the incorrect options. 
Our final set of problems includes 55 cases corresponding to eleven different conditions, which we use to sample 10 sequences of 30 problems each.

Table~\ref{tab:sample_explanations} contains examples of decision-making problems from both tasks.

\section{User Study Details}
\label{sec:appendix_userstudydetails}
\begin{figure*}[t]
    \centering
    \includegraphics[width=\linewidth]{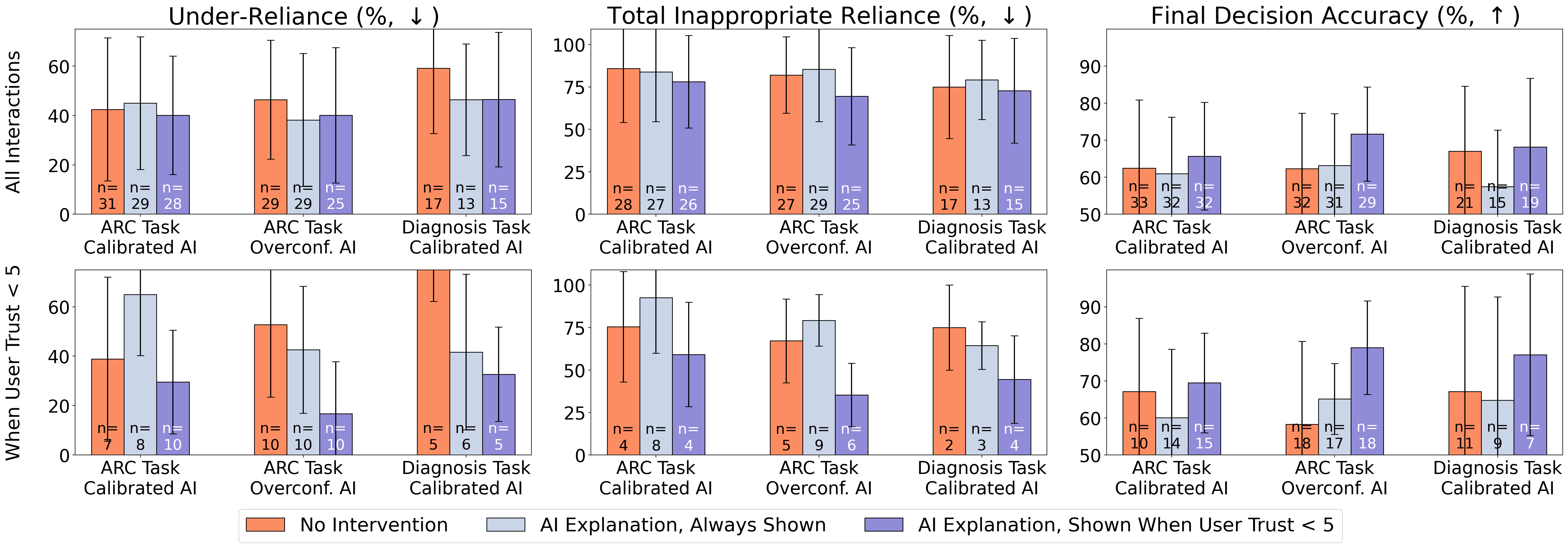}
    \includegraphics[width=\linewidth]{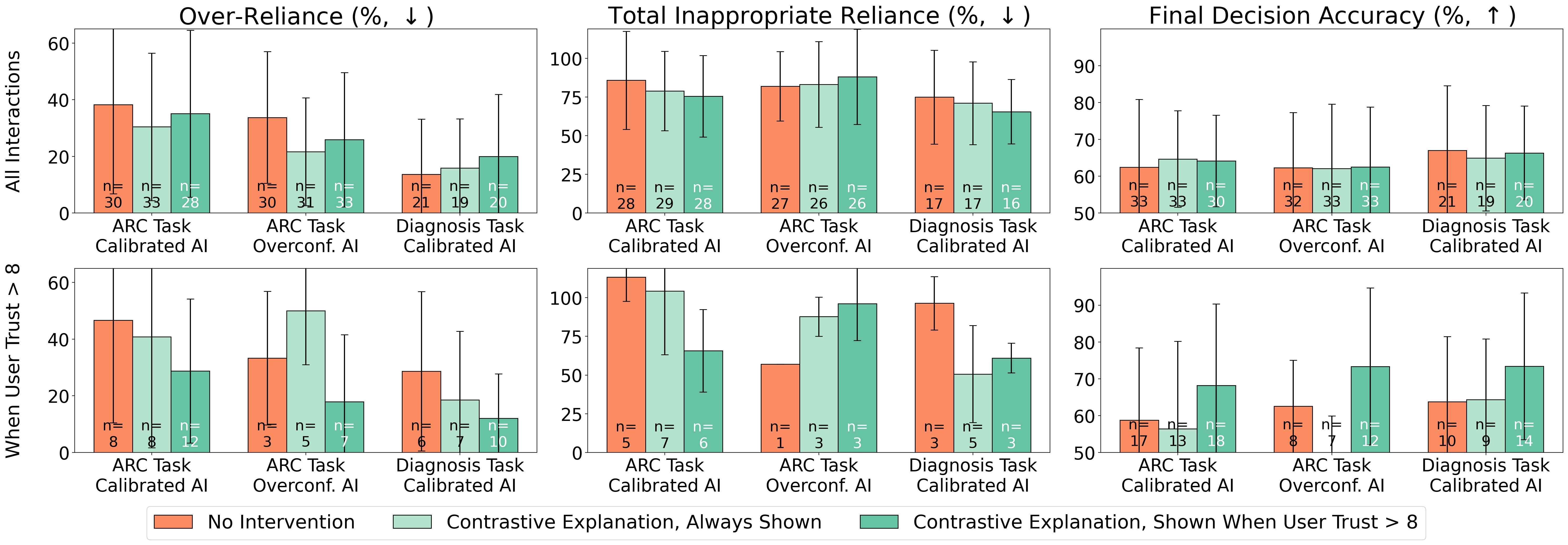}
    \caption{Macro-aggregation results for mitigating under- and over-reliance. $n$ represents the number of users who have at least 3 interactions meeting the required criteria.}
    \Description{Macro-aggregation results for mitigating under- and over-reliance. }
    \label{fig:macro-aggregation}
\end{figure*}

We present additional details about the user studies.

\paragraph{User Payment.} As mentioned earlier, users in the ARC task are paid a base payment of \$1.0, with an incentive of \$0.10 per correct answer. 
Users achieved 65--75\% accuracy on the task, which translates to a bonus of $\approx \$2.0$ per user, or \$3.0 total payment. 
The tasks took a median time of 15 minutes to complete, which translates to a pay rate of \$12 per hour. 

For the ARC task, users are paid a higher base payment of \$2.0, since the task takes slightly longer to complete ($\approx 20$ minutes). 
Users achieve 75\% accuracy, translating to an average bonus of \$2.0 per user, or \$4.0 total payment, which translates to a pay rate of \$12 per hour.

In total, including Prolific fees, we spent $\approx$ \$2500 on the user studies reported in this paper.

\paragraph{User Demographics.} Users were recruited on the Prolific platform. We recruited participants from the U.K. and U.S.A. who self-identified as fluent in English, and had at least 99\% approval rate on previously completed Prolific studies. 

\paragraph{Instructions.} Users were informed that they were participating in a research study. Figure~\ref{fig:interface} contains screenshots of the task instructions and payment details presented to users.
To maintain user attention, participants were instructed not to change their browser tab during the study, and a pop-up notification would send a warning whenever they did change their tab. To further ensure that users read the questions and choices closely before making their initial decision, the buttons for selecting the initial decision are only made active after 10 seconds.

Our user study was approved by our institution's Institutional Review Board (camera-ready version will include the name of the institution).

\section{Macro-Aggregation Results}
\label{sec:appendix_macroaggregation}

We report macro-aggregation results, where we first compute our metrics for each user, and then aggregate across all users. 
Because some users may only have a few interactions that fit the required criteria for computing metrics, the metrics for those users are very sensitive to a single interactions. 
Therefore, when computing metrics, we filter out users with fewer than 3 interactions that meet the required criteria ($\aidecision_i \neq \userdecision{\init}_i$, and user trust being below/above the corresponding threshold for the ``low trust''/``high trust'' interaction subsets). 
For the \underreliance\ metric we also need at least 3 interactions where the AI prediction is correct, and for the \overreliance\ metric we need at least 3 interactions where the AI prediction is incorrect.
For the \totalinapprel\ metric we \emph{both} need 3 interactions where the AI prediction is correct and 3 interactions where the AI is incorrect (so that we can calculate both \underreliance\ and \overreliance\ for that user).

In Figure~\ref{fig:macro-aggregation}, we observe trends similar to those observed in Section~\ref{sec:mitigations}. However, when looking at the low/high trust subsets, most conditions have fewer than 10 users for calculating \underreliance\ and \overreliance, and fewer than 5 users for \totalinapprel.


\section{Testing for Confounding Variables}
\label{sec:appendix_confoundingvariables}
\begin{table}[h]
    \centering
    \small
    \begin{tabular}{p{0.04\textwidth} p{0.08\textwidth} p{0.06\textwidth} p{0.08\textwidth} p{0.08\textwidth}}
    \toprule
        Users \newline Report \newline Trust? & \finalacc\ & \switchrate\ & \texttt{Under-\newline Reliance} & \texttt{Over-\newline Reliance}  \\
        \midrule
         \checkmark & 61.9\% & 48.6\% & 40.7\% & 34.8\% \\
          $\times$ & 60.2\% & 49.9\% & 42.0\% & 37.2\% \\
        \bottomrule
    \end{tabular}
    \caption{Effect of asking users to report trust on reliance metrics and decision accuracy, on the \arcc\ task setting.}
    \label{tab:trustreporting}
\end{table}

We evaluate the effect of asking users to report trust on their decision-making performance and reliance on the AI assistant, finding that user performance stays largely the same (Table~\ref{tab:trustreporting}).

\begin{table}[t]
    \centering
    \small
    \begin{tabular}{p{0.18\textwidth} p{0.05\textwidth} p{0.06\textwidth} p{0.06\textwidth}}
    \toprule
        How often user uses AI & \switchrate\ & \texttt{Under-\newline Reliance} & \texttt{Over-\newline Reliance}  \\
        \midrule
         Once a week or more & 43.4\% & 43.5\% & 29.9\% \\
          About once a month & 41.7\% & 41.4\% & 28.9\% \\
          Never & 41.8\% & 44.8\% & 26.9\% \\
        \bottomrule
    \end{tabular}
    \caption{Effect of user experience with AI on reliance metrics, on the \arcc\ task setting.}
    \label{tab:userfamiliarity}
\end{table}
\begin{table}[t]
    \centering
    \small
    \begin{tabular}{p{0.2\textwidth} p{0.05\textwidth} p{0.06\textwidth} p{0.06\textwidth}}
    \toprule
        How often user uses AI  \newline for checking facts & \switchrate\ & \texttt{Under-\newline Reliance} & \texttt{Over-\newline Reliance}  \\
        \midrule
         Regularly & 41.9\% & 46.7\% & 30.1\% \\
          Have used, not regularly & 45.4\% & 39.3\% & 29.9\% \\
        \bottomrule
    \end{tabular}
    \caption{Effect of user experience with AI for fact checking on reliance metrics, on the \arcc\ task setting.}
    \label{tab:userusage}
\end{table}

Table~\ref{tab:userfamiliarity} and Table~\ref{tab:userusage} show how user experience with AI, and specifically using it for fact checking, affects user reliance on AI assistance, again finding that reliance behavior is similar.
\clearpage

\begin{table*}[ht]
    \centering
    \small
    \begin{tabular}{p{0.43\textwidth} p{0.52\textwidth}}
         \toprule
         Supporting explanation generation prompt & Counter-explanation generation prompt \\
         \midrule
        Provide a convincing 3-4 sentence explanation (under 100 words) for why option \{$i$\} is the correct diagnosis, even if you don't believe option \{$i$\} is the correct diagnosis. It is absolutely essential that you justify that option \{$i$\} is the correct diagnosis, rather than any of the other options. \newline Do not explicitly refer to any options by their option number, only their content. For example, if you are explaining why option 1: Anemia is correct, do not say ``Option 1 is correct because...'' or ``For the above reasons, the correct diagnosis is option 1''), instead say ``Anemia is the correct diagnosis because...''. & Provide a convincing 3-4 sentence explanation (under 100 words) for why option $\{i\}$ may not be the correct diagnosis, even if you believe that option $\{i\}$ is the correct diagnosis. Your explanation should reflect that you believe option $\{i\}$ is correct (e.g. by saying ``While I think <option $\{i\}$> is the correct diagnosis...''), but also provide 1-2 reason(s) for why it may not be correct.  Your explanation should not merely restate the option, but provide a (short) reasoning. Do not explicitly refer to any options by their option number, only their content. For example, if you are explaining why option 1: Anemia is correct, do not say ``Option 1 is correct because...'' or ``For the above reasons, the correct diagnosis is option 1''), instead say ``Anemia is the correct diagnosis because...''. \\
         \bottomrule
    \end{tabular}
    \caption{Prompts used for generating supporting and counter-explanations for the Diagnosis task. Prompts for ARC task were slightly modified (e.g. replacing ``correct diagnosis'' with ``correct answer'').}
    \label{tab:prompts}
\end{table*}

\begin{table*}[ht]
    \centering
    \small
    \begin{tabular}{p{0.06\textwidth} p{0.24\textwidth}  p{0.1\textwidth} p{0.3\textwidth} p{0.2\textwidth}}
         \toprule
         Task & Question & Options & Supporting Explanation & Counter-Explanation \\
         \midrule
         ARC & Water has many unique chemical properties. Which property of water makes water a good solvent of crystalline salts? & Low Ph \newline \textbf{Strong polarity} & Water's strong polarity allows it to interact with the charged ions in crystalline salts, breaking them apart and dissolving them. This is because water molecules are polar, with oxygen being more electronegative than hydrogen, creating a partial negative charge near the oxygen and partial positive charges near the hydrogen atoms. This polarity allows water molecules to surround and solvate the ions in the salt, leading to the dissolution of the salt in water. & Although strong polarity is the primary reason water is a good solvent, low pH could potentially enhance the solubility of certain crystalline salts by increasing the availability of hydrogen ions to interact with the salt molecules. \\
         \midrule 
         Diagnosis & Patient is a 7 year old male.\newline- The patient lives with 4 or more people.\newline- The patient attends or works in a daycare.\newline- The patient has pain somewhere, related to their reason for consulting.\newline- The patient's pain is sensitive.\newline- The patient feels pain in the right tonsil, under the jaw, and in the trachea.\newline- The intensity of the patient's pain is 3 (on a scale of 0 to 10).\newline- The patient's pain does not radiate to another location.\newline- The patient's pain appeared at a speed of 2 (on a scale of 0 to 10).\newline- The patient has a fever, either felt or measured with a thermometer.\newline- The patient has had a cold in the last 2 weeks.\newline- The patient has noticed that the tone of their voice has become deeper, softer, or hoarse. &- Epiglottitis \newline - Viral pharyngitis \newline - \textbf{Acute laryngitis} \newline - Chagas & Acute laryngitis is the correct diagnosis because the patient exhibits hoarseness and a change in voice tone, which are hallmark symptoms of laryngitis. The presence of a recent cold and mild pain in the throat area align with viral-induced laryngitis, typically following upper respiratory infections. The patient's fever and pain in the right tonsil and trachea further support inflammation consistent with laryngitis. & While I think acute laryngitis is the correct diagnosis, it may not be entirely accurate because the patient's pain is specifically located in the right tonsil, under the jaw, and in the trachea, which might also suggest a more localized infection like tonsillitis or pharyngitis. Additionally, acute laryngitis typically involves a more pronounced hoarseness of voice, which may not align perfectly with the patient's description of a deeper or softer tone, potentially pointing towards other conditions. \\
         \bottomrule
    \end{tabular}
    \caption{Examples of supporting and counter-explanations generated by GPT-4o. The AI prediction (which is also the correct option) is highlighted.}
    \label{tab:sample_explanations}
\end{table*}

\end{document}